\pdfoutput=1

\documentclass[useAMS,usenatbib]{mn2e}

\usepackage{graphicx}
\usepackage{ulem}
\usepackage{subfigure}
\usepackage{lscape}
\usepackage{rotating}

\def\degree{\ifmmode {^\circ}\else {$^\circ$}\fi}
\def\rstar{\ifmmode {\, R_{\star}}\else $R_{\star}$\fi}
\def\msol{\ifmmode {\, M_{\odot}}\else $M_{\odot}$\fi}
\def\rsol{\ifmmode {\, R_{\odot}}\else $R_{\odot}$\fi}
\def\lsol{\ifmmode {\, L_{\odot}}\else $L_{\odot}$\fi}
\def\msolyr{\ifmmode {\, M_{\odot}\,{\rm yr}^{-1}}\else $M_{\odot}\,{\rm yr}^{-1}$\fi}
\def\mdot{\ifmmode {\,\dot{M}}\else $\dot{M}$\fi}
\def\mdotyr{\ifmmode {\,\dot{M}\,yr^{-1}}\else $\dot{M}\,yr^{-1}$\fi}

\newcommand{\Teff}{\ifmmode{T_{\rm eff}}\else{$T_{\rm eff}$}}

\title[VLTI observations of the dust geometry around RCB stars]
  {\bf{VLTI observations of the dust geometry around R  Coronae Borealis stars}\thanks{Based on observations made with the Very Large Telescope Interferometer at Paranal Observatory under program 079.D-0415.}}
\author[S.N. Bright et al.]
  {S.N.~Bright,$^1$
  O.~Chesneau,$^2$ G.C. Clayton,$^3$ O. De Marco,$^1$
  I.C. Le\~ao,$^4$ 
    \newauthor % starts a new line in the
J. Nordhaus,$^5$
  J. S. Gallagher$^3$
  \\
  $^1$Department of Physics \& Astronomy, Macquarie University, Sydney, NSW 2109, Australia
     \\
  $^2$UMR 6525 H. Fizeau, Univ. Nice Sophia Antipolis, CNRS, Observatoire de la C\^{o}te d'Azur, Av. Copernic, F-06130 Grasse, France
 \\
 $^3$Department of Physics \& Astronomy, Louisiana State University, Baton Rouge, LA 70803, USA
\\
$^4$Departamento de F\'isica, Universidade Federal do Rio Grande do Norte, 59072-970 Natal, RN, Brazil
\\
 $^5$Department of Astrophysical Sciences,  Princeton University, Princeton, NJ 08544, USA}
\date{Received, Accepted}
\def\LaTeX{L\kern-.36em\raise.3ex\hbox{a}\kern-.15em
    T\kern-.1667em\lower.7ex\hbox{E}\kern-.125emX}

\begin{document}

\label{firstpage}

\maketitle

\begin{abstract}
We are investigating the formation and evolution of dust around the hydrogen-deficient supergiants known as R Coronae Borealis  (RCB) stars.  We aim to determine the  connection between the probable merger past of these stars and their current dust-production activities
 
We carried out high-angular resolution interferometric observations of three RCB stars, namely RY Sgr, V CrA, and V854~Cen with                                                                                                                                                                                                                                                                                                                                                                                                                                                                                                                                                                                                                                                                                                                                                                                                                                                                                                                                                                                                                                                                the mid-IR interferometer, MIDI on the VLTI, using two telescope pairs. The baselines ranged from 30 to 60 m, allowing us to probe the dusty environment at very small spatial scales ($\sim$ 50 mas or  400 \rstar).  The observations of the RCB star dust environments were interpreted using both geometrical models and one-dimensional radiative transfer codes.

From our analysis we find that asymmetric circumstellar material is apparent in RY Sgr, may also exist in V CrA, and is possible for V854 Cen.  Overall, we find that our observations are consistent with dust forming in clumps ejected randomly around the RCB star so that over time they create a spherically symmetric distribution of dust.  However, we conclude that the determination of whether there is a preferred plane of dust ejection must wait until a time series of observations are obtained.  
% This guide is for authors who are preparing papers for
% \textit{Monthly Notices of the Royal Astronomical Society} using the
%\LaTeXe\ document preparation system and the {\tt mn2e} class file.
\end{abstract}

\begin{keywords}
RY Sgr, V Cra, V854 Cen
          Techniques: interferometric; Techniques: high angular
               resolution; Stars: circumstellar matter; Stars: mass-loss
\end{keywords}

\section{Introduction}
\label{sec:intro}
The R Coronae Borealis (RCB) stars are a small group of carbon-rich supergiants. Their defining characteristics are hydrogen deficiency and unusual variability \citep{1996PASP..108..225C}. RCB stars undergo massive declines of up to 8 mag due to the formation of carbon dust at irregular intervals. Two scenarios \citep{1996ApJ...456..750I,2002MNRAS.333..121S} have been proposed for the origin of RCB stars: the first is the merger of a helium and a carbon-oxygen white dwarfs that temporarily results in a swollen star \citep{1984ApJ...277..355W}, while the second is a final helium-shell flash taking place on a white dwarf that temporarily increases the star's radius to supergiant dimensions \citep{1977PASJ...29..331F,1979ASSL...75..155R}. Final-flash stars have been directly observed. In the case of Sakurai's object, which suffered an outburst in 1995, the entire evolution from compact hot star to supergiant star with RCB variability was monitored with modern telescopes, creating a convincing link between final-flash and RCB stars. However, recently \citet{2005ApJ...623L.141C,2007ApJ...662.1220C} made the discovery that RCB stars have $^{18}$O/$^{16}$O ratios that are orders of magnitude higher than those seen in any other known stars, including Sakurai's object  \citep{2002Ap&SS.279...39G}. These high ratios are inconsistent with regular stellar evolution, but can be produced in the aftermath of a merger \citep{2007ApJ...662.1220C,2010ApJ...714..144G}. This finding tips the balance of evidence in favour of a merger origin for RCB stars.

RCB stars have effective temperatures in the range 5000-7000~K so circumstellar gas in thermodynamic equilibrium would not be able to condense into dust at distances closer than $\sim$20 \rstar. However, we observe a correlation between pulsational phase and the onset of dust formation in five RCB stars which indicates that dust must condense close to the photosphere \citep{2007MNRAS.375..301C,1999AJ....117.3007L,1977IBVS.1277....1P}. It is possible that shocks created by the puslations
will cause local density enhancements, and encourage non-equilibrium
conditions. Then the preconditions for carbon
nucleation may be temporarily present  \citep{1992A&A...265..216G,1996A&A...313..217W}.
Such conditions existing locally over the surface of the star could allow a ``clump''
of carbon dust as close as 2 \rstar to form which is then ejected by radiation pressure.  The observed time scales for
RCB dust formation fit  well with those calculated by carbon chemistry models \citep{1986ASSL..128..151F,1996A&A...313..217W}.

Even though all RCB stars show an infrared (IR) excess, there is no increase in the excess seen at the time of a major decline in brightness.  This indicates that the amount of new dust formed in one dust formation episode is small compared to the total mass emitting around the star \citep[and references therein]{1996PASP..108..225C,1997MNRAS.285..317F}. 

Spectro-polarimetry of the prototype RCB star, R CrB, obtained during a deep decline showed %optical depth changes as a function of wavelength that are 
a change in the position angle of the continuum polarisation consistent with a bipolar geometry, which includes a thick disk or torus that obscures the star and additional diffuse dust above the poles \citep{1997ApJ...476..870C}. Large resolved shells with sizes ranging from 5\arcsec~to 20\arcmin~have been seen around various RCB stars at visible and IR wavelengths showing shapes ranging from spherically symmetric to slightly elliptical \citep{1986ApJ...310..842G, 1985A&A...152...58W,1986ASSL..128..407W, 1999ApJ...517L.143C, 2001ApJ...560..986C,ClaytonPerCom2010}.

Using near-IR interferometry, \citet{2003A&A...408..553O} found evidence for an asymmetrical dust geometry  around R CrB at $\sim$20 \rstar from the star in addition to a thin dust shell at 60-80 \rstar. \citet{2004A&A...428L..13D} found an asymmetrical arrangement of dust clouds around RY Sgr in the $K$- and $L$-bands using adaptive optics.  These dust clouds were found at $\sim$0$\farcs$1-0$\farcs$2 from the star corresponding to 700-1400 \rstar. Asymmetrical dust was also seen at 100 $\rstar$ around RY~Sgr in 2005 using interferometry in the mid-IR  \citep{2007A&A...466L...1L}.

In this paper we investigate the immediate circumstellar environments of three RCB stars, RY~Sgr, V854~Cen and V~CrA, with the aim of determining a connection between the probable merger past of these stars and their current dust-production activities. In \S\ref{sec:obs} we describe our observations; in \S\ref{sec:modeling} we report our modelling procedures. In \S\ref{sec:Results} we provide the results of the modelling of the three sources, and finally, in \S\ref{sec:discuss} we discuss  our results in the context of past work.

%The standard format for papers submitted to \textit{Monthly
%Notices} is \LaTeXe. The layout design for \textit{Monthly
%Notices} has been implemented as a \LaTeXe\ class file. The {\tt
%mn2e} class file is based on the {\tt mn} style file, which in
%turn is based on \verb"article" style as discussed in the \LaTeX\
%manual \citep{la}. Commands that differ from the standard \LaTeX\
%interface, or that are provided in addition to the standard
%interface, are explained in this guide. This guide is not a
%substitute for the \LaTeX\ manual itself. We also refer authors to
%\citet{kd} and \citet{kn}. Authors planning to submit their papers
%in \LaTeX\ are advised to use \verb"mn2e.cls" as early as possible
%in the creation of their files.

\section{Observations}
\label{sec:obs}

\begin{table*}

\begin{caption}
{VLTI/MIDI Observation Log
\label{tab:VLTIlog}}
\end{caption}

\begin{tabular}{lllccc}\hline
{\small Observation}   & {\small Date}  & {\small Baseline}   & {\small Label} &\multicolumn{2}{c}{{\small Projected baseline}}  \\
(g=grism)& &  & & Length  & PA   \\
(p=prism)&& & &{\small(meters)} & {\small (degrees)} \\
\hline
RY Sgr-g & 2005-06-25 & U1-U4 & B$_1$ & 122 & 34 \\
RY Sgr-g & 2005-06-25 & U1-U4 &B$_2$ & 123 & 36 \\
RY Sgr-g &  2005-06-25 & U1-U4 &B$_3$ & 128 & 65 \\
RY Sgr-g &  2005-06-26 & U1-U4 & B$_4$ &125 & 68 \\
RY Sgr-g &  2005-05-26 & U3-U4 &B$_5$ &  57 & 98 \\
RY Sgr-g &  2005-06-28 & U3-U4 &B$_6$ & 62 & 110 \\
\smallskip
RY Sgr-g &  2005-06-26 & U3-U4 &B$_7$ & 57 & 135 \\
RY Sgr-g & 2007-06-29 & U2 - U3 &B$_1$ &  47 & 32  \\
RY Sgr-g  & 2007-06-29 & U2 - U3 & B$_2$ & 41 & 54 \\
RY Sgr-g  &  2007-06-29 & U2 - U3 & B$_3$ &34 & 58  \\
RY Sgr-g  &  2007-06-30 & U3 - U4 &B$_4$ & 55 & 94  \\
\smallskip
RY Sgr-g  &  2007-06-30 & U3 - U4 &B$_5$ & 59 & 129  \\
V CrA-p  &  2007-06-29 & U2 - U3 & B$_1$ & 45 & 45  \\
V CrA-p  &  2007-06-29 & U2 - U3 & B$_2$ & 32 & 63  \\
V CrA-p  &  2007-06-30 & U3 - U4 & B$_3$ & 53 & 89  \\
\smallskip
V CrA-p  &  2007-06-30 & U3 - U4 & B$_4$ & 61 & 128  \\
V854 Cen-g  &  2007-06-30 & U3 - U4 & B$_1$ & 60 & 101  \\
V854 Cen-p  &  2007-06-30 & U3 - U4 & B$_2$ & 56 & 153  \\
\hline
\end{tabular}

{\tiny Results of the RY Sgr observations taken in 2005 were published by \citet{2007A&A...466L...1L}.

}
\end{table*}

\begin{table*}

\begin{caption}
{ISO and Spitzer Observation Log
\label{tab:IRlog}}
\end{caption}

\begin{tabular}{lcc}\hline 
Star & Instrument/Telescope & Date   \\
\hline
RYSgr & SWS/ISO & 1997-03-25 \\
%\hline
V CrA & IRS/Spitzer  &  2005-09-14  \\ 
%\hline
V854 Cen & SWS/ISO  & 1996-09-09    \\ 
%\hline
V854 Cen & IRS/Spitzer  &  2008-06-12  \\ 
\hline

\end{tabular}

{\tiny }
\end{table*}

{ The sources were observed in June 2007 with the Very Large Telescope Interferometer (VLTI) MID-infrared Interferometric instrument (MIDI) \citep{2003Ap&SS.286...73L, 2007A&A...471..173R}. 
% In addition, we used archival MIDI observations of RY\,Sgr reported by Le\~ao et al. (2007). 
The VLTI/MIDI Interferometer operates like a classical Michelson interferometer combining the mid-IR light (N band, 7.5-13.5 $\mu$m) from two VLT Unit Telescopes (UTs, 8.2 m). Eleven observations were completed in June 2007 using two different telescope pairs (U2-U3 and U3-U4).  A typical MIDI observing sequence was followed, as described in \citet{2007A&A...471..173R}. 
MIDI provided single-dish acquisition images with a spatial resolution of about 250 mas at 8.7$\mu$m, flux-calibrated spectra at low and high spectral resolution (R=25, 230 respectively), and visibility curves. %Visibility curves are the Fourier  transform of the light along a baseline. 
The spatial spectrum of the source, or visibility function, is given by the two-dimensional Fourier transform of the sky brightness.  The interferometric information along a baseline is identical to the one-dimensional Fourier transform of the curve resulting from the integration of the brightness distribution in the direction perpendicular to the baseline. The observations were performed in High-Sens mode, implying that the photometry of the sources is recorded subsequent to the fringes. We used both the grism ($\lambda$/$\Delta$$\lambda$ = 230) and prism ($\lambda$/$\Delta$$\lambda$ = 25) for wavelength dispersion.  %The following calibrators were used in our observations: HD152334 K4III 3.99$\pm$0.07\,mas, HD163376 M0III 3.79$\pm$0.12\,mas, HD169916 K1III 3.75$\pm$0.04\,mas, HD177716 K1III 3.72$\pm$0.07\,mas. 
See Table \ref{tab:VLTIlog} for additional details.  

The image acquisition field-of-view is 3\arcsec. A ``chopped" image is produced by tilting the secondary mirror off-centre and taking an image of the sky followed by an image being taken at the centre position.  This process of tilting back and forth is done rapidly several times.  The resulting images are subtracted to cancel the sky background leaving an image with only the source signal. %is taken after the coarse acquisition by the telescopes. 
This image is used to
position the sources to a predetermined pixel in order to maximise
the overlap of both images for the interferometric
measurements. The MIDI images have a clear Airy pattern and the full width at half maximum (FWHM) is consistent with the source being unresolved at 8.7 $\mu$m, with an angular diameter $<$ 150-200 mas, given the error bars.  In absence of this nodding technique the quality of the image is limited and the dynamics are restricted to about 2 to 3\% of the flux at peak. %However, the dynamics are very limited as the restrictive field of acquisition is about 3 arcsec. %\textbf{?Olivier, what do you mean by this last sentence?}

Next, the beam combiner and the dispersive device are inserted, producing two interferometric
beams of opposite sign. The zero optical path
difference (OPD) is searched for by scanning around the expected
point of path length equalisation. Once the OPD is found the interferometric measurements start using
self-fringe-tracking. A temporal fringe pattern is produced by scanning constantly at a range of
OPD = 40-80 $\mu$m typically in steps of 2 $\mu$m.

We used two different MIDI data reduction packages: MIA
developed at the Max-Planck-Institut f\"ur Astronomie and EWS
developed at the Leiden Observatory (MIA+EWS\footnote{Available at http://www.strw.leidenuniv.nl/$\sim$nevec/MIDI/index.html}, V.1.5.1). %Errors in the visibilities range from 8\% to 15\%. 
MIDI spectra were calibrated using the stars HD152334 (K4III 3.99$\pm$0.07\,mas), HD163376 (M0III 3.79$\pm$0.12\,mas), HD169916 (K1III 3.75$\pm$0.04\,mas), and HD177716 (K1III 3.72$\pm$0.07\,mas). The accuracy of the absolute flux calibration is better than 10\%. The observation log is given in Table \ref{tab:VLTIlog}.  

As the photometry of individual telescopes is recorded a few minutes after the fringe recording, the fluctuations of the atmosphere affect the absolute level of the visibilities at the 8-15\% level (depending on the atmospheric conditions). This is by far the worst source of error in this mode.   Because the knowledge of the level of the fluctuations is lacking, it is not possible to easily decrease this error. Furthermore, the error affects  all the disperser channels almost identically, and are hence highly correlated to the atmospheric error. There are also other sources of noise that affect the spectral channels independently at a level 3-5 times lower than the atmospheric error.  These include the photon noise and the mismatches of the beam overlap effect. This implies that the differential information contained in the slope of the dispersed visibility curve is more tightly constrained than the absolute level. This mix of correlated and uncorrelated errors make any model fitting using the classical approach of the reduced $\chi^2$ ill-suited. However, our reduced $\chi^2$ remains an appropriate  indicator of the fit at first approximation and will be used to compare several models.

%Images in the mid-IR of the target V854\,Cen were also secured using the instrument VISIR at the VLT during the night of the 1st of July 2008. These observations were made with 3 filters: PAH1 (8.59$\mu$m, half bandwidth 0.42$\mu$m ), 
%SiC (11.85$\mu$m, 2.34$\mu$m) and NeII (12.81$\mu$m, 0.21$\mu$m).
%We used the imager in burst mode, using a pixscale of 0.075 arcsec and a field of view of 
%19.2 $\times$ 19.2 arcsec. With the burst mode, all the singles chopping and nodding images 
%are recorded, allowing the reconstruction of quality-enhanced images using shift and add 
%techniques. We used the standard chopping/nodding technique to remove the background, with a perpenticular chop throw, 
%a chopping frequency of 0.25 Hz and an amplitude of 8 arcsec. 
%We shifted and added the images using a maximum of correlation algorithm, after removing the bad images, selected as the one for which the measured
%flux was smaller than the mean flux of all the images minus one sigma.
%The great observing conditions during our run (0.43 mm of water in the atmosphere) allowed us to obtain great quality 
%diffraction-limited images. The images were flux calibrated and deconvolved using standard stars nearby V854Cen as a flux template and as a measure of the Point Spread Function (PSF).  A maximum of likelihood Richardson-Lucy algorithm with $\sim$50 iterations was used.

Differential phase measurements  are  provided by MIDI and  are well-suited to detect source asymmetry, however it has limitations.   
The phase extracted is the departure of the spectral phase compared to the expected phase of the fringes given that the OPD is known. As a differential measure, the absolute phase is not recovered in the data reduction process. Moreover, the slope of the differential phase is also removed. As a consequence, the differential phase is mostly sensitive to complex objects composed of at least two sources exhibiting very different spectral signature within the N band, or sources with large variations of the phases, such as well resolved binary sources \citep[e.g.][]{2009A&A...502..623R}.  As such, it is particularly well suited for the study of bright stellar sources surrounded by dusty disks \citep[e.g.][] {2007A&A...467.1093D, 2008A&A...490..173O}.
We did not detect such large amplitude phase signal for the two reasons presented above, namely,  the spectral changes of the differential phase induced by the presence of the clumps are too slow and strongly decreased in the reduction process and the spectral signature of the clumps do not differ significantly enough to induce a large spectral signature that would provide a definite signature.  Therefore, we are not able to use the differential phases for direct proof of asymmetries in the circumstellar material around the RCB stars.

Spitzer Infrared Spectrograph (IRS)  Short-low (SL) and Long-low (LL) resolution spectra of V854 Cen and V CrA were obtained from the Spitzer archive. An existing IRS spectrum of RY Sgr is heavily saturated and was not used in our analysis. The V854 Cen spectrum was obtained on 12 August 2008 and the V CrA spectrum was obtained on 14 September  2005. The spectra were re-reduced using the SMART software package as described in \citet{2006ApJS..165..568F}.
Infrared Space Observatory (ISO) spectra taken with the Short Wavelength Spectrometer (SWS) were obtained for RY Sgr (25 March 1997) and for V854 Cen (9 September 1996) \citep{2001ApJ...555..925L}. These data were obtained from the ISO archive and were re-reduced using a slightly modified version of the SWS routines developed by \citet{2003ApJS..147..379S}. A low S/N ISO spectrum of V CrA is present in the archive but was not used in our analysis. An observation log of the ISO and Spitzer spectra is found in Table \ref{tab:IRlog}. Light curves using the American Association of Variable Star Observers (AAVSO) and All Sky Automated Survey (ASAS) photometry for the three RCB stars are presented in \S\ref{sec:discuss}.

\begin{figure*}

  \centering
  \subfigure[ ] { 
  \includegraphics* [width=4.cm]  {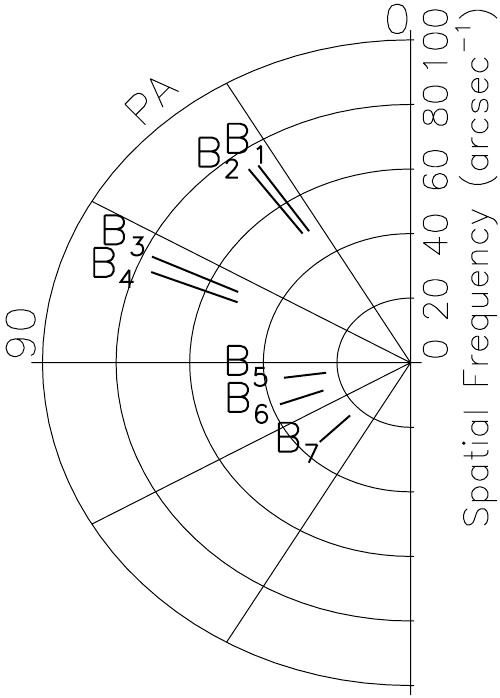}
  }
  \subfigure[ ] { 
  \includegraphics* [width=4.cm]  {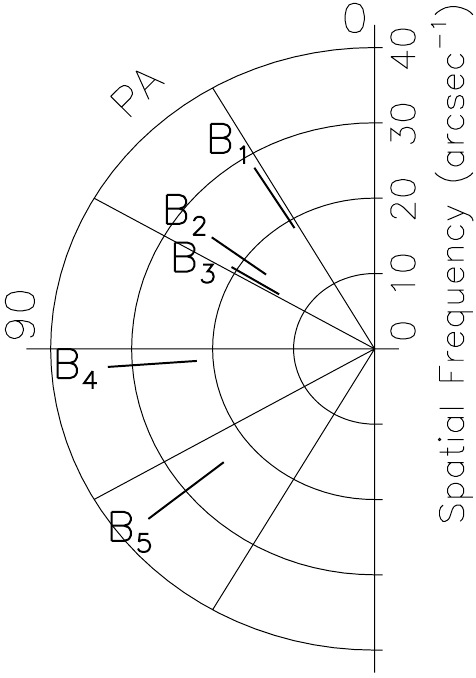}
  }
  \subfigure[ ] { 
  \includegraphics[width=4.cm] {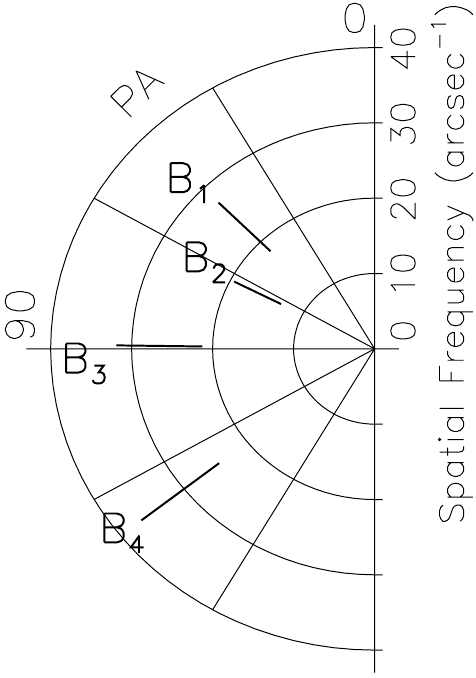}
        }
  \subfigure[ ] { 
  \includegraphics[width=4.cm] {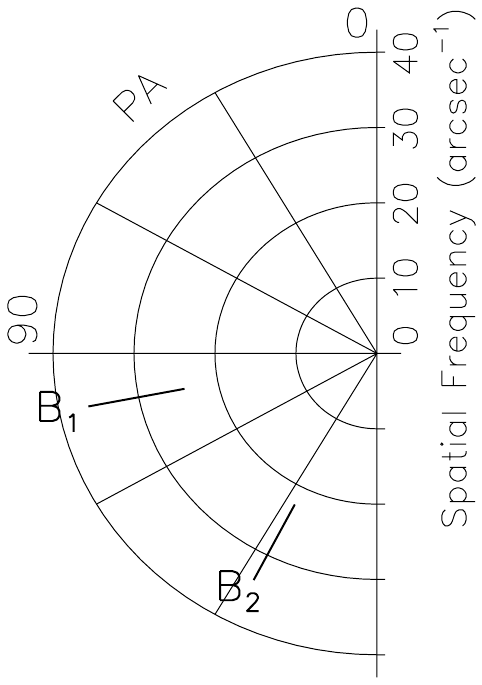}
        }

     %\label{fig_vis1}
     \caption{Charts of the \textit{uv}-plane for (a) RY Sgr in 2005, (b) RY Sgr in 2007 (c) V CrA  and (d) V854 Cen are shown, representing their projected baseline lengths and position angles (PAs) with the same labels as in Table 1. Note the increased spatial frequency range for RY Sgr in 2005 data compared to the 2007 observations.
     \label{fig:uvplane}}

   \end{figure*}

\section{Modelling of the dusty environments}
\label{sec:modeling}
\begin{table*}

\begin{caption}
{Parameters of the monochromatic geometric fits to the visibility curves.  The models are as follows: PS: Point Source; UD: Uniform disk; G: Gaussian Shell (representing a spherical distribution of clumps that are individually unresolved);  C: Cluster (an off-centred point source representing  an asymmetry in the distribution of dust clumps with an ambiguity of 180\degree) and EUD: Elliptical uniform disk (with different major and minor axis)
 \label{tab:geoMod}}
\end{caption}
\begin{tabular}{llccccccccc}\hline 
Star   & Model  & $\chi 2$ & Flux PS  & Flux Shell/Disk  & FWHM$^a$  & Diam Disk$^b$ & Flux Cluster& Sep of Cluster & P.A.$^c$  & \\
   (epoch) &   &  & (\%) & (\%) & (mas) &(mas) & (\%)  & (mas) & ($^\circ$) &  \\ 

\hline
RY Sgr (2005)  & PS +  UD       & 1.7 & 0.21 & 0.79 & - &33& -   & -  & - &   \\
RY Sgr (2005)  & PS + G + C &    0.31 & 0.11 & 0.8 & 18 &-& 0.09& 15& 79 &  \\
\smallskip
RY Sgr (2005)  & PS + EUD & 1.1 & 0.2 & 0.8 &- &18, 13& - & -& 79 &\\
RY  Sgr (2007)  & PS + G         & 14 & 0.05 &0.95 & 36 &   - & -   &  -    & -   &   \\
RY Sgr (2007) & PS + UD        & 12 &0.17  & 0.83    & - &46& -  & -  & -   &\\
RY Sgr (2007)  & PS + G + C & 1.3 & 0.15 & 0.74 & 37& -& 0.11 & 25 & 175 & \\
RY Sgr (2007)  & PS + UD + C & 1.8 & 0.19 & 0.63 & - &58& 0.18 & 23 & 172  &\\
\smallskip
RY Sgr (2007)  & PS + EUD & 2.7 & 0.17 & 0.83 & - &36, 19&- & - & 178  &\\
%\hline
V CrA  & PS + G         & 0.81  & 0.56   & 0.44 & 46 & - &  - & -   & -   & \\

V CrA  & PS + UD        & 0.73    & 0.60 & 0.4 & -& 70  &   -& -  & -   & \\
V CrA  & PS + G + C & 0.26    & 0.55  & 0.4& 45  &-& 0.05 & 84& 16 &  \\
V CrA  & PS + UD + C & 0.27 & 0.61  & 0.34 &- &76	& 0.05	& 100& 9 &  \\
V CrA  & PS + EUD (a)$^d$ & 0.41 & 0.58 & 0.42 &-& 180, 52 & - & - & 166 &\\
\smallskip
V CrA  & PS + EUD (b)$^d$ & 0.61 & 0.59 & 0.41 & - &96, 63& - &-  & 177 &\\
%\hline
V854 Cen  & PS + G & 0.58  & 0.16 &0.84  &24  & -& -& - &-& \\
V854 Cen  & PS + G + C & 0.8 & 0.3  & 0.6 & 24 &-& 0.1 & 33 & 40  & \\

\hline
\end{tabular}

{\tiny $^a$FWHM of Gaussian Shell 

$^b$ Diameter of UD or major axis, minor axis of EUD

$^c$P.A. of cluster or orientation of EUD

$^d$Two PS + EUD models for V CrA with different major and minor axes that produce similar results   }
\end{table*}

%Extensive modeling is needed to understand the RCB environments as we have a very limited sampling of the \textit{uv}-plane. 
%In interferometry, data are taken are sample ``cuts" across the source.  
Extensive modeling is needed to understand the RCB star environments. Model parameters can be better constrained if many  observations across the \textit{uv}-plane are obtained.  Observations were taken across different baselines and position angles (PA) for each star as seen in Figure \ref{fig:uvplane}   and a few  visibility measurements for each object were obtained.   %Following in suite, if an entire sampling of the \textit{uv}-plane (across all PAs and baselines) was obtained a complete image of the star and its environment could be made.  
%Because of the limitation of VLTI/MIDI (a few hours of observing only provides a few points of visibility) a large sampling of the \textit{uv}-plane is unlikely. Therefore, extensive modeling is needed to understand the RCB environments  

Two approaches are used in order to model the data. First, following the approach published by \citet{2007A&A...466L...1L}, we use several geometric models to account for the dispersed visibilities, first obtaining a monochromatic fit and then seeking for wavelength dependent (chromatic) solution.  The geometrical models provide information on departure from spherical symmetry in the circumstellar environment.  
Second, we use the 1D radiative transfer code, {\tt DUSTY}, a public-domain simulation code that models radiation transport in a circumstellar dusty environment.  {\tt DUSTY} analytically integrates the radiative-transfer equation in plane-parallel or spherical geometries \citep{1999astro.ph.10475I,1997MNRAS.287..799I}. {\tt DUSTY} provides a realistic view of the source, as it uses both the dispersed visibilities and the spectral energy distribution (SED) of the source. However, {\tt DUSTY} does not provide any information on departure from spherical symmetry.

The two modelling approaches are complementary, provided that one insures broad agreement in the areas of overlap. For our models, described in detail in \S\ref{sec:Results}, this agreement is sought between the geometrical and {\tt DUSTY} models of the shells only.

\begin{table}

\begin{caption}
{RCB Target Stars; basic characteristics%data and calculated observational resolution
\label{tab:basic}}
\end{caption}
%\begin{tabular}{|llcccccc|}\hline 
%Star & RA ($^\circ$ $'$ $''$) & M$_V$ &A$_V$* & V* & T$_{eff}$* & D**   & Resol.\\
%& Dec (h m s) & (mag) &(mag) & (mag)  & (K) & (pc)  & (mas/AU)    \\ 
%\hline
%RYSgr & 19 16 33 & -5 &0.06 & 6.4 &  7250 &1800 & 0.55 \\ &  -33 31 20 & & & &  & \\
%\hline
%V CrA & 18 47 32 & -4 &0.37 & 9.95  &  6250 & 5200  &  0.19 \\ &  -38 09 32 & & &  & & \\
%\hline
%V854 Cen & 14 34 49 &-5& 0.00 & 7.1 & 6750& 2600 &   0.38  \\ &  -39 33 19 & &  & & &  \\

%\hline

%\end{tabular}

\begin{tabular}{llcccc}\hline 
Star & M$_V$$^a$ &A$_V$$^b$ & V$^b$ & T$_{eff}$$^c$ & D  \\ %& Spatial Resolution\\
&  (mag) &(mag) & (mag)  & (K) & (pc) \\ %& (mas/AU)    \\ 
\hline
RY Sgr & -5 &0.06 & 6.4 &  7250 &1800 \\ %& 0.55\\
%\hline
V CrA & -4 & 0.37 & 9.9  &  6250 & 5200  \\%&  0.19 \\ 
%\hline
V854 Cen & -4.5 &  0.00 & 7.1 & 6750& 2100 \\%&   0.48  \\ 
\hline

\end{tabular}

{\tiny 
$^a$\citet{2001ApJ...554..298A, 2009A&A...501..985T}

$^b$From \citet{1990MNRAS.247...91L}

$^c$From \citet{2000A&A...353..287A} }
\end{table}

\begin{table*}

\begin{caption}
{ DUSTY model inputs and outputs
\label{tab:dustyMod}}
\end{caption}
\renewcommand{\tabcolsep}{0.27cm}
%\renewcommand{\arraystretch}{2}
%\begin{tabular}{|l|c|c|c|c|c|c|c|c|c|c|c|c|c|}  \hline
\begin{tabular}{l c c c c c c c c  c c c c }  \hline

\multicolumn{1}{c}{} & \multicolumn{8}{l}{Input Parameters } &  \multicolumn{4}{l} {Output Results}\\ \hline

%Star &\begin{sideways} $T_{eff}$ ($K$) \end{sideways}&\begin{sideways} Dust density power exponent ($y^{-P}$) $ $\end{sideways}&\begin{sideways} T$_{dust}$  at $R_{inner}$ ($K$) \end{sideways}&\begin{sideways}$\tau_V$ \end{sideways}& \begin{sideways} Shell Thickness \end{sideways} & \begin{sideways}Star's Distance ($pc$) \end{sideways} & \begin{sideways} Stellar Luminosity ($L_\odot$)\end{sideways} & \begin{sideways}E(B-V)\end{sideways} & \begin{sideways} \rstar ($mas$) \end{sideways}&\begin{sideways} $R_*$ ($R_\odot$) \end{sideways}& \begin{sideways} Shell  $R_{inner}$ (mas) \end{sideways} & \begin{sideways} Shell $R_{inner}$ ($R_*$) \end{sideways}& \begin{sideways} $\chi^2$ of Visibility Fit \end{sideways}\\ 

Star & $T_{eff}$ & $P$ $^a$ & $T_{inner}$$^b$ & $\tau_V$ & \begin{math} \frac {R_{out}}{R_{in}} \end{math} $^c$ & D & L & E(B-V) & \rstar  & $R_{in}$$^d$   & $R_{{\tt DUSTY}}$$^e$  &$\chi^2$ $^f$ \\ 
& ($K$) & & ($K$) & & & ($pc$) & ($L_\odot$) &  & ($R_\odot$) & ($mas$), (\rstar) & ($mas$), (\rstar)&  \\

\hline
RY Sgr 2005 &	7250 &3.3 & 900	 & 1.2 &50 & 1800 &	8700 &	0.02	& 59 & 7, 48 & 13, 79 &1.9\\ %\hline
RY Sgr 2007	&7250&2.7&1000&1.0&	50&1800 &8700	&0.02 & 59 & 9, 57 & 17, 108 &0.6\\ %\hline
V CrA & 6250 & 1.7   & 1000 &	1.5&	30	 & 5200&	4500 &	0.12  &  57 & 2, 41 &5, 100 &1.3\\ %\hline
V854 Cen & 6750 &	2.7 &1075 &0.45&50&2100&8200&	0&	66 & 6, 38 & 11, 71 &4.3\\ \hline

\end{tabular}
{\tiny 

$^a$Dust density is $r^{-P}$;     $^b$T$_{dust}$  at R$_{inner}$;  $^c$  Shell thickness in multiples of R$_{inner}$;      

  $^d$Shell inner radius;         $^e$HWHM of the {\tt DUSTY} geometrical Gaussian fit at 9 $\mu$m;           $^f$Reduced $\chi^2$ of visibility fits for a subset of baselines only (see text, \S\ref{sec:Results}). 
}
\end{table*}

\subsection{Geometrical Fitting}
\label{ssec:GeoFit}
%Closely following the procedure described in \citet{2007A&A...466L...1L} we extensively fitted the visibility curves from the MIDI/VLTI using different geometrical models.  Because of the limited sampling of the \textit{uv}-plane our models may be degenerate.  With more observations across the \textit{uv}-plane comes better constrained parameters and more accurate fits can be obtained, but ours was a limited sample. The   

If the circumstellar material is spherically symmetric each sampling with baselines of similar projected length, but different PA will produce the same visibility curve.  However, the data shows that this is not  the case despite the limited number of baselines considered, particularly in the case of RY Sgr. Therefore, models are used to interpret the visibility curves and determine the possible geometry of the circumstellar material.
In our geometrical models we consider different combinations of the following structures: a central point source representing the star,  a Gaussian shell of dust around the star (i.e. a circular structure with a Gaussian light distribution), a circular, uniform dusty disk around the star, an non-centrosymmetric, elliptical dusty disk around the star (because of the simplicity of our models this could represent either a circular disk observed at an angle of inclination or an actual elliptical disk observed face on), and a second point source representing a dust ``cluster" near the star.  
Typically, RCB stars are not thought to have homogenous dust shells, so in our models a ``shell'' represents an approximate spherical distribution of clumps that are individually unresolved.  The physical interpretation of the second point source (the cluster) is an estimate for the level of asymmetry in the distribution of dust clumps surrounding the RCB star.  %The parameters of our models are listed in Table \ref{tab:geoMod}.   
As our data were collected with telescope pairs, closure phase information was not obtained and therefore there is an ambiguity of 180\degree in the position of the cluster.

If one obtains observations that employ only short VLTI baselines, the \textit{uv}-plane coverage is limited.  Therefore, the fits from our geometric models are only indicative of what may be surrounding the RCB star, as the models are not unique. Nevertheless, the extent of model degeneracy was tested by extensive exploration of parameter space. 

%Following in suite, if an entire sampling of the \textit{uv}-plane (across all PAs and baselines) was  obtained we could make a complete image of the star and its environment.  However, because of the limitation of VLTI/MIDI (a few hours of observing will only give you a few points of visibility) a full data-set is unlikely.  

At the start of our fitting procedure, we neglect any variation of the morphology as a function of wavelength. Every combination of structures modelled results in a set of monochromatic visibility curves. This approach gives us a basic understanding of the global morphology of the object. The value of the reduced $\chi$$^2$ is minimised for all the baselines. Only the errors in the mean visibility level are considered. The parameters of the monochromatic models are listed in Table \ref{tab:geoMod} and results displayed in Figures \ref{fig:RY2007Geo}a, \ref{fig:VCrAGeo}, and \ref{fig:V854CenGeo}.   %The errors on shape (i.e. on the slope of the visibilities versus wavelength) are neglected because we are only interested in looking for a global monochromatic solution.   

Because the differential information contained in the slope of the dispersed visibility is more tightly constrained than the absolute level (see \S\ref{sec:obs}), our analysis relies heavily on the shapes of the dispersed visibility curves rather than the absolute level of the visibility curves.    Therefore it is not surprising that we obtain some reduced $\chi^2$  values less than 1 for our models.  %In this case, it may be suggested that the error bars have been overestimated.  However, the absolute level of the visibility curves does have a dramatic influence on the spatial scale of the dust shell. 
In this context, a reduced $\chi^2$ value of 0.2 remains better than a reduced $\chi^2$ value of 0.8 for the same model, as it means that the dust shell density distribution is better accounted for. %, since the shape of the dispersed visibility curve is a better match.

Next, a chromatic analysis of visibilities across all wavelengths is performed. The separation of the cluster from the star and its PA is assumed to be the same at all wavelengths. However, the FWHM of the shell and the flux of all components are wavelength dependent and are therefore adjusted to fit wavelength-dependent visibility curves.  Results of the  chromatic fits for RY Sgr are displayed in Figures \ref{fig:RY2007Geo}b and \ref{fig:RY2007Geo}c. % \ref{fig:VCrAGeo}b, \ref{fig:VCrAGeo}c, and \ref{fig:V854CenGeo}b,  \ref{fig:V854CenGeo}c.  
Fits were also completed for V854 Cen and V CrA, but are not presented in this paper as they did not provide adequate information.    
Specific results of the geometrical fitting for each RCB star are presented in \S\ref{sec:Results}.

%\textbf{ ?Olivier or Izan, do you have a better short description on the FWHM vs wavelength and Flux vs wavelength (the third panel in Figs. 2, 5, 7)?}

\subsection{ {\tt DUSTY}  Fitting}
\label{ssec:DustyFit}

{\tt DUSTY} utilises the self-similarity and scaling relations of the radiatively-heated dust, e.g. the shell is uniquely  characterised by its optical depth. This means that absolute values (synthetic surface brightness, flux, and visibility) are not uniquely determined by the transfer problem and must be inferred by external constraints - in our case %the recorded MIDI visibilities and flux-calibrated spectra, and 
the stellar luminosity and distance.

{\tt DUSTY} requires seven main input parameters: the temperature of the central star, the chemical composition of the dust grains, the grain size distribution, the density distribution of the dust, the temperature of the dust at the inner radius, the overall radial optical depth at a given wavelength, and the thickness of the dust shell.  {\tt DUSTY}'s  outputs consist of a detailed SED,  the detailed surface brightness at our specified wavelengths (8 - 13 $\mu$m), tables of the radial profiles of density, optical depth and dust temperature, and visibility as a function of the spatial frequency for the specified wavelengths. The synthetic visibility profiles throughout the N-band (7.5 - 13 $\mu$m) are generated using a set of 15 wavelengths.  These are compared with the observed MIDI visibilities for each baseline. 

Because {\tt DUSTY}  is a 1D radiative transfer code it is unable to model non-spherically symmetric, non-homogenous (clumpy) structures. As a result, the observed visibility curves that  present evidence for asymmetric geometric structures cannot be reproduced accurately by the {\tt DUSTY} models. On the other hand, since radiative transfer is taken into account, {\tt DUSTY} can model the characteristics of the circumstellar dust (such as producing an SED) in ways that the geometric models cannot.
The input parameters for the {\tt DUSTY} models are given below:

\begin{itemize}

%\item The dust shell is spherically symmetric and homogeneous (i.e. not clumpy).

\item The temperature and the spectrum of the central star are constrained by model fitting, assuming a simple black-body source. The effective temperatures are based on \citet{1998A&A...332..651A, 2000A&A...353..287A}.

\item The dust is made entirely of amorphous carbon. This is the common assumption for these hydrogen-deficient stars \citep{2001ApJ...555..925L}. {\tt DUSTY} uses standard  optical constants for amorphous carbon from \citet{1988ioch.rept...22H}.

\item An MRN size distribution \citep{1977ApJ...217..425M} is assumed: $ n(a) \propto a^{-q} \quad \hbox{for} \quad a_{\rm min} \le a \le a_{\rm max}$.  For our analysis  we used  $q$=3.5 and varied $a_{\rm min}$ and $a_{\rm max}$ (see \S\ref{ssec:Grain}).%the values $q$=3.5, $a_{\rm min}$= 0.005 $\mu$m, and $a_{\rm max}$=2 $\mu$m are used.  

\item Inside the shell, the density follows an $r^{-P}$ distribution as in a steady-state wind with constant velocity.  The power exponent, $P$,  was the easiest parameter to constrain and as such was normally the first parameter chosen when fitting the SED and visibility curves.  A larger value causes most of the dust grains to concentrate near the inner radius ($R_{inner}$).  When grains are concentrated near $R_{inner}$ they absorb a lot of photons causing a ``screen" to form (this is especially apparent with smaller grains which absorb more efficiently).  The extent of this screen dramatically affects the slope of the visibility curve.  Therefore, a larger value of $P$ typically yields a steeper slope in the modelled visibility curves.      %It is important to note that the slopes of the visibility curves could not be matched for RY Sgr and V854 Cen without smaller grains, because with the larger grains, the screening effect could not occur.  

\item The temperature of the dust at the inner rim of the shell ($T_{inner}$) is chosen individually for each star and mainly constrained by the SED.  A higher temperature chosen here causes a larger infrared emission.%more flux in the infrared.

\item The optical depth, $\tau_V$, is chosen so as to match the V-band (0.55 $\mu$m) photometric measurement of the star. % fit each star at  V (0.55 $\mu$m).  
Like $T_{inner}$, a higher $\tau_V$ causes more flux in the infrared.  %In addition, $\tau_V$ is directly related to the density of dust within the shell, and hence the dust's brightness dictates the shape of the visibility curves.  A higher dust density in the shell yields a higher visibility and vice versa.

\item The thickness of the dust shell ($R_{out}/R_{in}$) is constrained mainly by the visibility curve but does not have a large effect. Once a certain thickness is reached the output visibility curves do not change noticeably until a thickness ratio of $\sim$ 90-100 is reached, at which point the object becomes over-resolved. Therefore, the smallest shell thickness that gives the best result is chosen, but this parameter is poorly constrained.
\end{itemize}

 In order to ease the comparison between the shell produced by {\tt DUSTY} models and those produced by the geometrical models a Gaussian curve is fit to the 1D collapsed intensity distributions of the best {\tt DUSTY} model at a chosen wavelength (9 $\mu$m).  %The agreement is good between the two different models except for V CrA for which a different solution is clearly favoured.  % A Gaussian is fit to the shell after the best fit {\tt DUSTY} model is determined.  The FWHM of this Gaussian fit is compared to the FWHM of the shells in the geometrical models.  
The half width at half maximum (HWHM) of the {\tt DUSTY} Gaussian fit (from here on called $R_{{\tt DUSTY}}$) is used as the distance to the shell in our analysis to make comparison easier. %in \S{\ref{sec:discuss}}. 

%The dust was set to be 100\% Amorphous Carbon with a standard MRN distribution for all stars. The power is chosen to fall off  for each star as y$^{-P}$. The outer radius of the dust shell is forced by the thickness constraint in DUSTY (example: DUSTY input Shell Thickness of 50 $\rightarrow$ R$_{outer}$ = 50 times R$_{inner}$).  

As discussed, {\tt DUSTY}  does not take into account the distance nor the luminosity of the star, so the user must scale  {\tt DUSTY}  results to  specific stars.  
%A short IDL program was written to better display and understand the results given by  {\tt DUSTY} .    Distance and stellar luminosity are inputs into the IDL code. 
The distances are calculated from the M$_V$, A$_V$, and V values in Table \ref{tab:basic} and the luminosity inputs are kept within the range of RCB stars  with these parameters.   %\textbf {\sout{ Is this an okay way to fit?  I tried to keep it within reasonable bounds, but are there any luminosity standards for these stars that I should use ???}}
The results are also corrected for  the standard CCM  reddening law \citep{1989ApJ...345..245C} using an input of $E(B-V)$, where $E(B-V)$=A$_V$/3.1.
The scaling process results in a  stellar radius and inner shell radius in physical units.   The inputs and outputs, including the reduced $\chi^2$, are displayed in Table \ref{tab:dustyMod}.  Specific results of the   {\tt DUSTY} fitting for individual stars are discussed in \S\ref{sec:Results}.

  \section{Results }
   \label{sec:Results}
\subsection{RY\,Sgr}
\label{ssec:RYResults}

\begin{figure*}

  \centering
   \subfigure[ ] {
  \includegraphics[width=4.725cm]{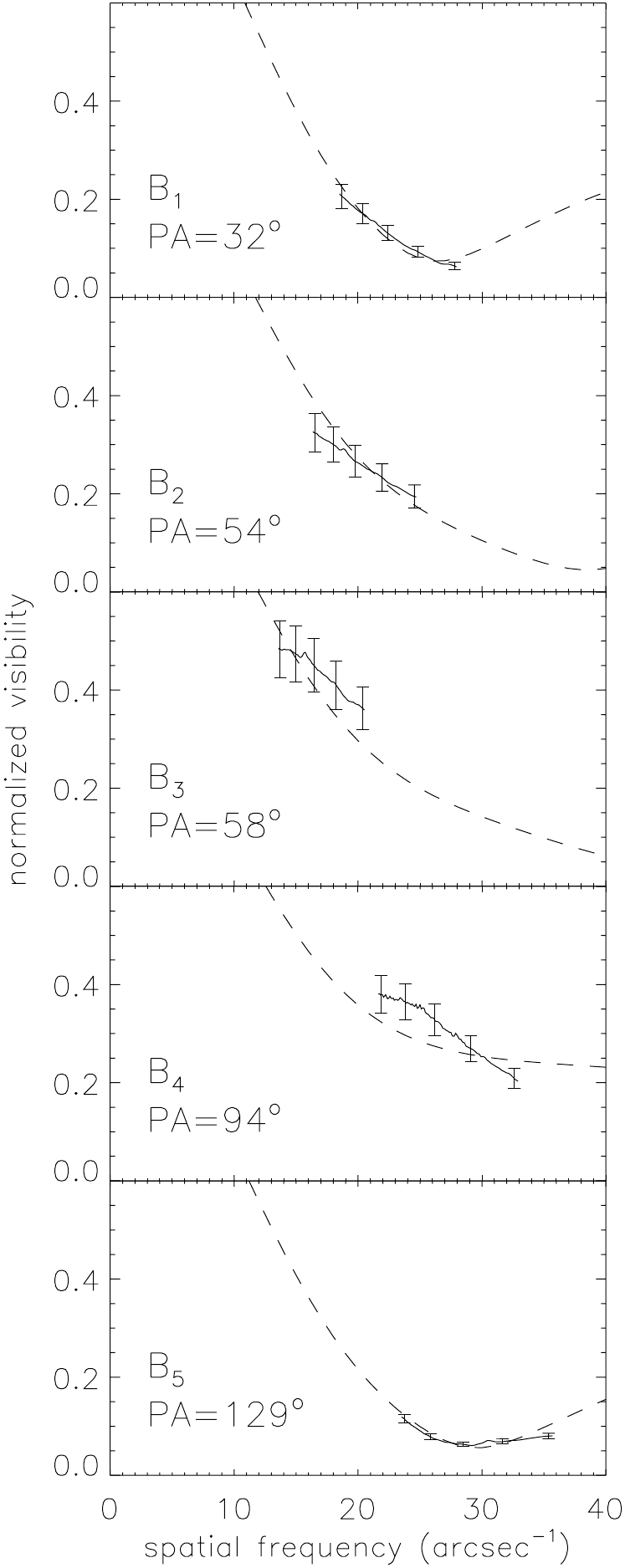}
  }
   \subfigure[ ] {
  \includegraphics[width=4.cm]{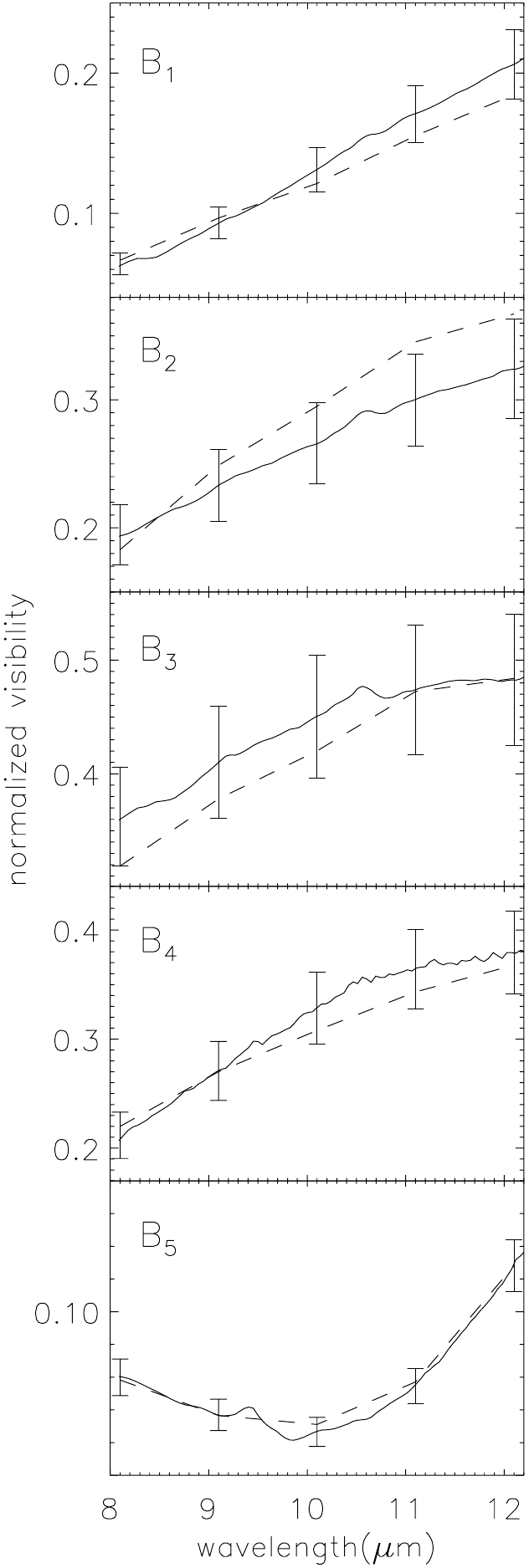}
  }
   \subfigure[ ] {
  \includegraphics[width=4.24cm]{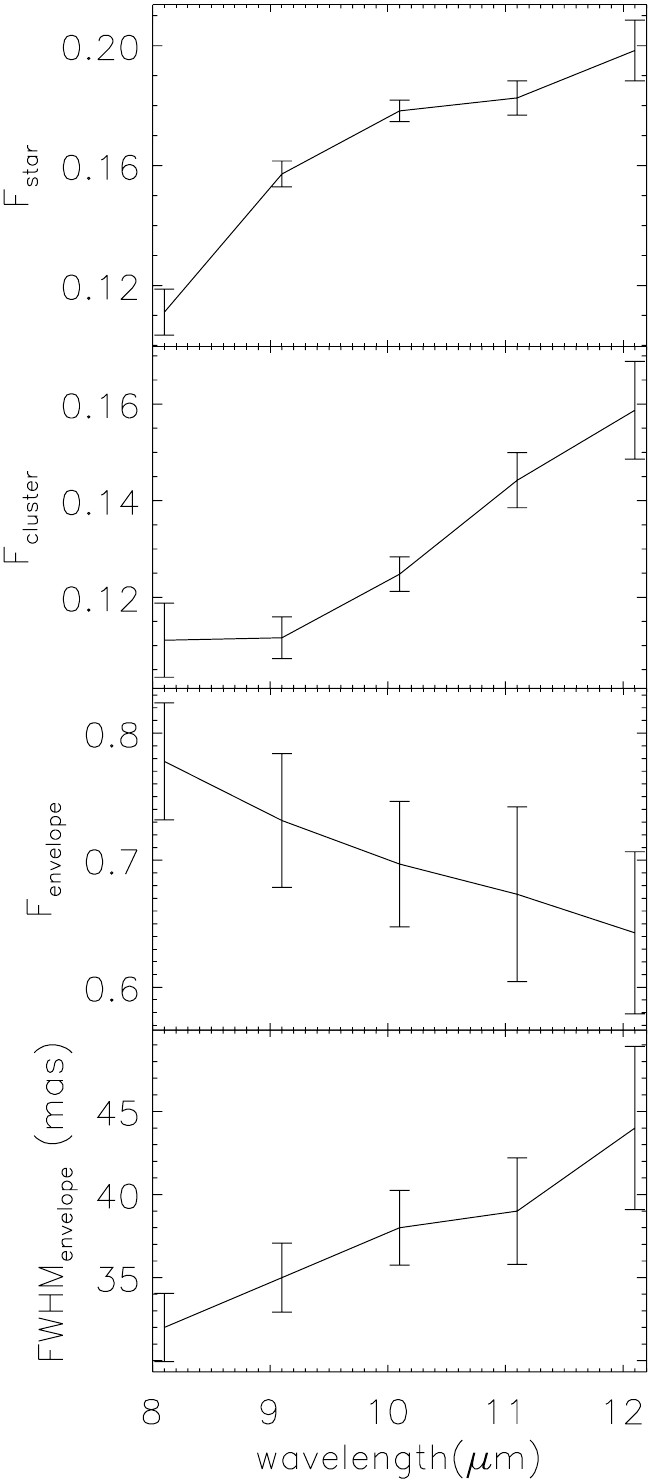}
  }
 
  \caption{Geometrical fits (dashed line) for RY Sgr in 2007 are shown using a model consisting of a point source, a Gaussian shell and a cluster. (a) The fits using the the monochromatic models are shown.  (b) Fits using the chromatic models are shown.  (c) Corresponding chromatic quantities derived from the best-fitting model. Each panel is labelled by the baseline and PA referenced in Table \ref{tab:VLTIlog} and Figure \ref{fig:uvplane}. The observed visibility curves (solid lines) have indicative error bars marked.  
    \label{fig:RY2007Geo}}

   \end{figure*}

\begin{figure*}

\begin{center}
    \subfigure[ ] {
  \includegraphics[width=5.65cm]{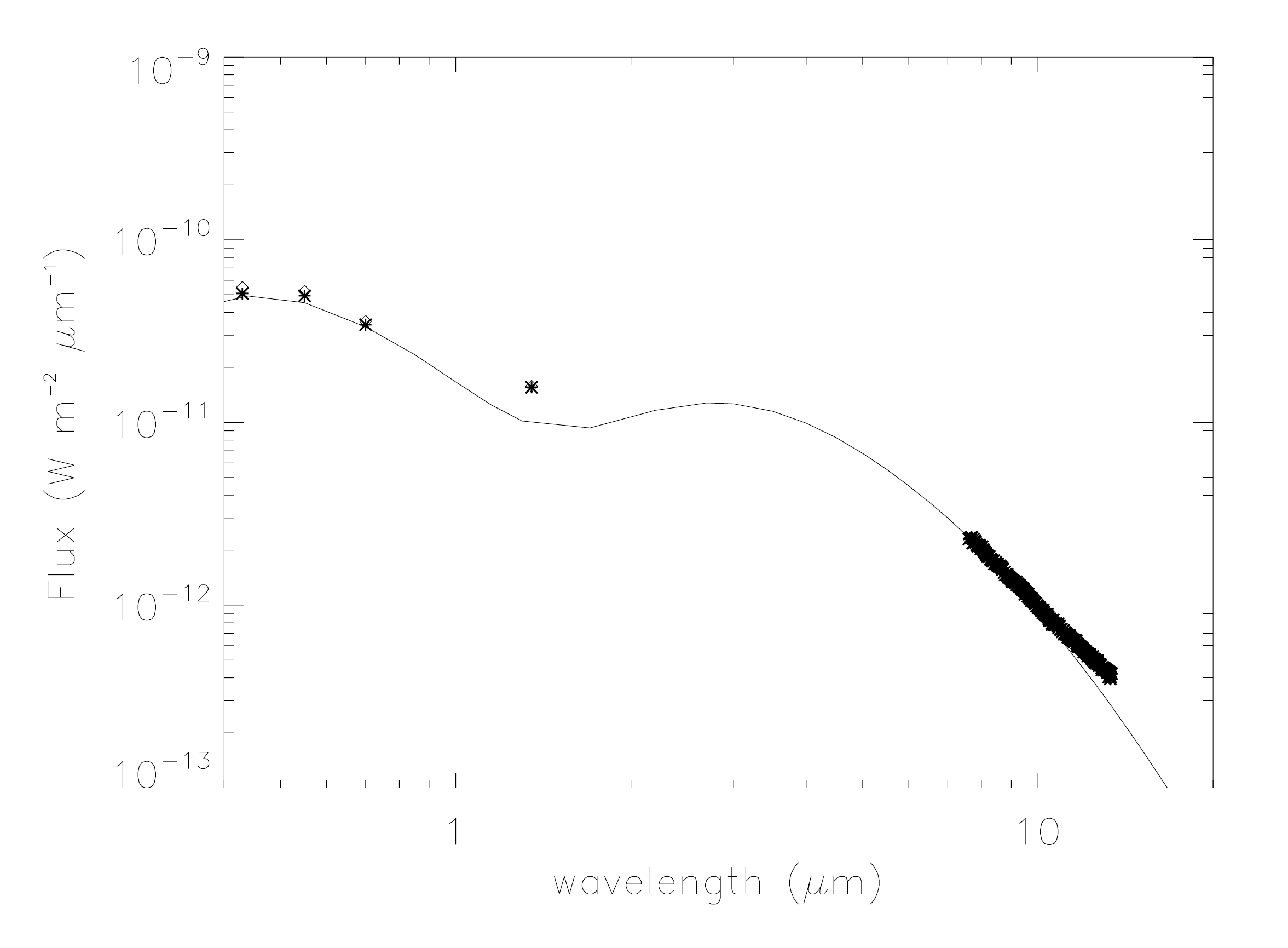}%VCrAfitmono_A.pdf}
  }
   \subfigure[ ] {
  \includegraphics[width=5.5cm]{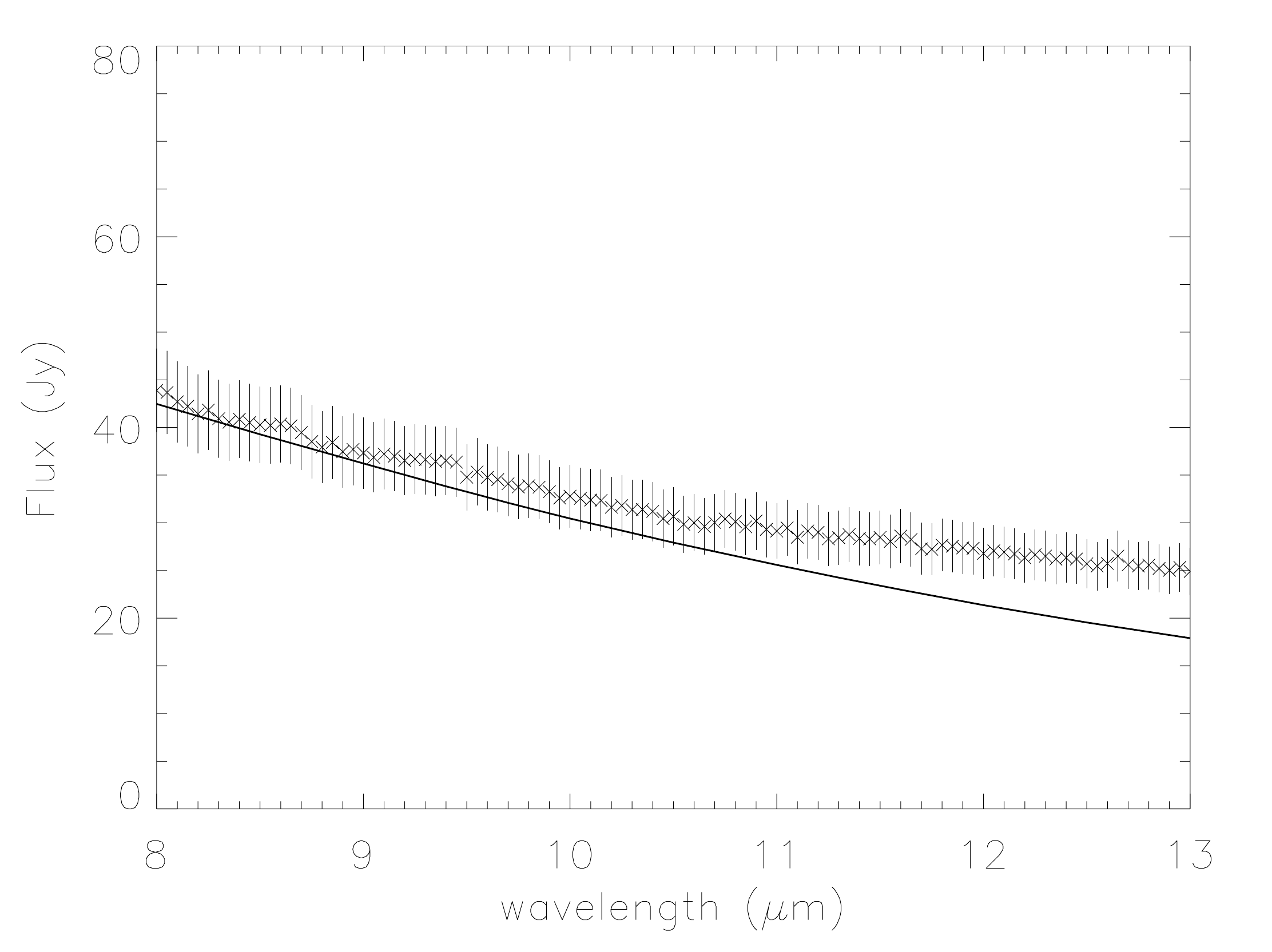}%VCrAvcurves_wl_A.pdf}
  }
   \subfigure[ ] {
  \includegraphics[width=5.6cm]{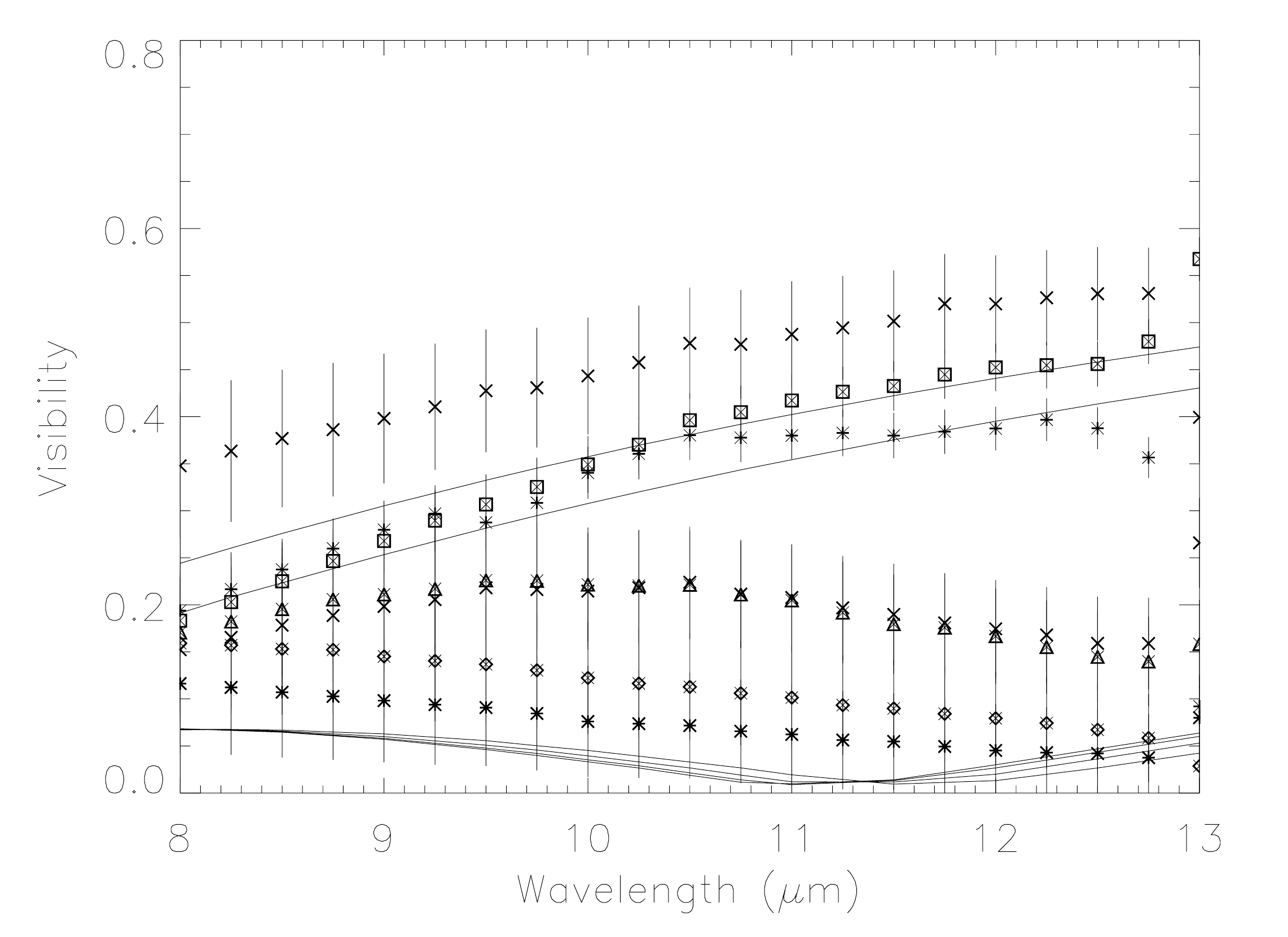}%VCrAparams_wl_A.pdf}
}
     \caption{Best {\tt DUSTY} fits for RY Sgr in 2005.  (a) The modelled SED (solid line) is shown along with the observed spectra (symbols). (b) Only the MIDI spectrum  (crosses) is shown with the model (solid line). (c) The 7 baseline visibility curves (symbols) are shown with the visibility fits (solid line). The visibility curves are shown from the shortest baseline (top) to the longest baseline (bottom). Note only B$_5$ and B$_6$ can be fit.  The other baselines, B$_1$, B$_2$, B$_3$, B$_4$, B$_7$, show strong evidence of asymmetry which {\tt DUSTY}  cannot account for (see \S\ref{sssec:RYGeoFit}).   Indicative error bars are marked.   \label{fig:RY2005Dusty}}
  \end{center}	
   \end{figure*}

 \begin{figure*}
      
\begin{center}
 % \centering
  %\subfigure[ ] { 
   \subfigure[ ] {
  \includegraphics[width=5.85cm]{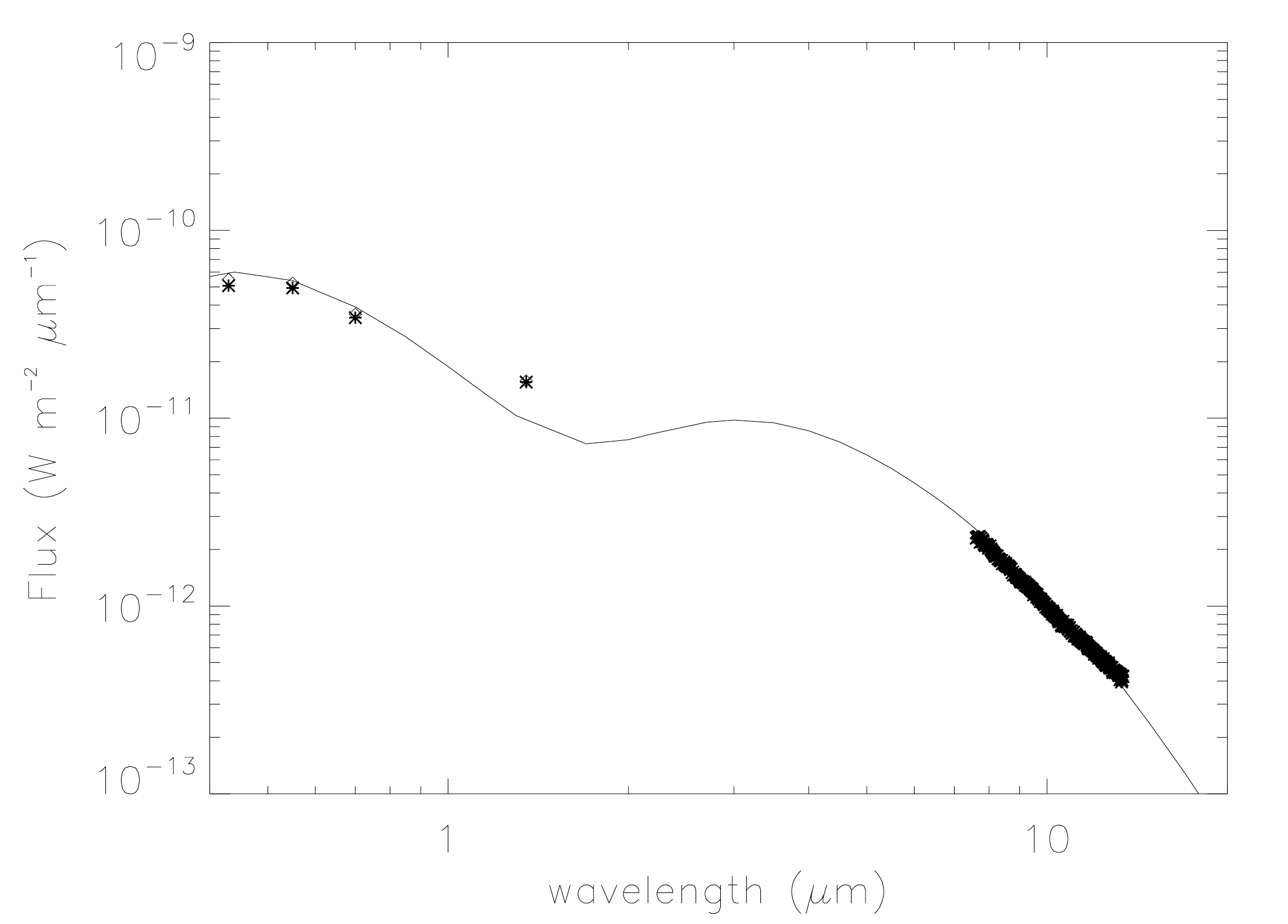}%VCrAfitmono_A.pdf}
  }
   \subfigure[ ] {
  \includegraphics[width=5.5cm]{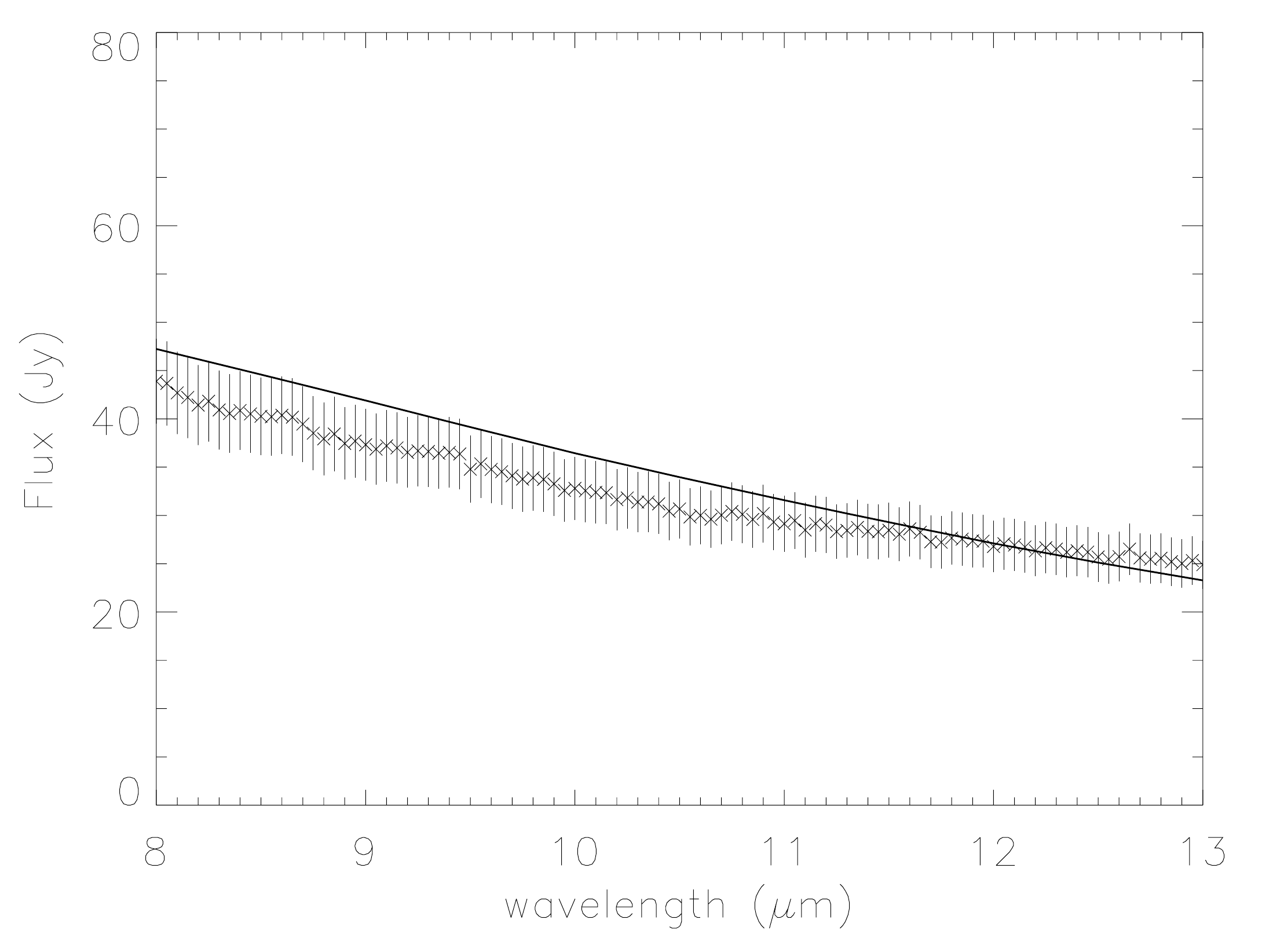}%VCrAvcurves_wl_A.pdf}
  }
   \subfigure[ ] {
  \includegraphics[width=5.7cm]{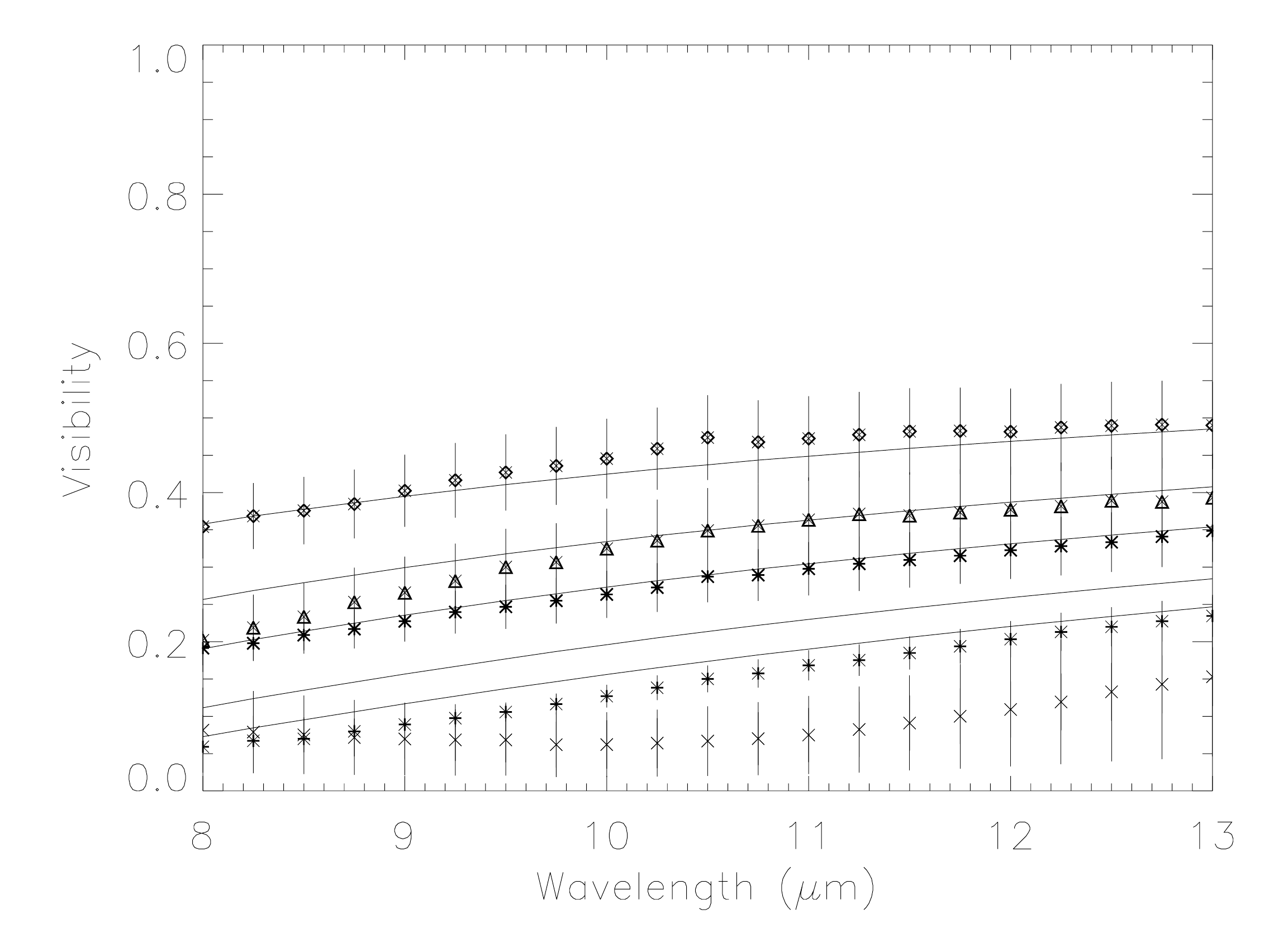}%VCrAparams_wl_A.pdf}
  }
%%\includegraphics[width=7.8cm]{rysgr_comp_2005-2007.pdf}
%  \includegraphics [width=7.cm] {RY2007SED.pdf}\\
%  %}
%  %\subfigure[ ] {
%  \includegraphics[width=7.cm] {RY2007Flux.pdf}\\
%    %    }
%  %\subfigure[ ] { 
%  \includegraphics[width=7.cm] {RY2007VIS.pdf}
%      %  }

     %\label{fig_vis1}
     \caption{Best {\tt DUSTY} fits for RY Sgr in 2007.  (a) The modelled SED (solid line) is shown along with the observed spectra (symbols). (b) Only the MIDI spectrum  (crosses) is shown with model (solid line) (c) The 5 baseline visibility curves (symbols) are shown with the visibility fits (solid line). The visibility curves are shown from the shortest baseline (top) to the longest baseline (bottom). Note only the three longer baselines (B$_1$, B$_2$, B$_3$) can be fitted (see \S\ref{sssec:RYGeoFit})   Indicative error bars are marked.    \label{RY2007Dusty}}  
 \end{center}	
   \end{figure*}

 \subsubsection{Geometrical Fitting}
   \label{sssec:RYGeoFit}

Using the geometrical models described in \S\ref{ssec:GeoFit} we find that the best monochromatic model (with the lowest reduced $\chi^2$) for the RY Sgr 2007 dataset is a combination of a central star, a Gaussian dust shell and a dust cluster (see Figure \ref{fig:RY2007Geo}a).

A significant wavelength dependence of the model parameters
was found for the 2007 dataset (see Figure \ref{fig:RY2007Geo}c). The FWHM
of the envelope grows with increasing wavelength and is accompanied
by a flux increase of both the star and cluster. Such a behaviour is counter-intuitive.  In depth investigations of the degeneracy between the flux and size parameters were performed by first, forcing the flux of the envelope to increase with wavelength and second, by allowing the cluster position to be free. After a large exploration of the parameter space, a good parameter set was not found. Next, a more simple geometrical model, involving a stellar unresolved source and a Gaussian dust shell, was tested. In this case, the flux of the envelope increased with wavelength, but the
fit was not satisfactory (reduced $\chi^2$=12). To summarise, the strongly suspected asymmetries in the 2007 data can be ascribed, at first order, by the presence of a cluster, however the model has some problems. ``Clusters" introduce a spectrally dependent sinusoidal pattern in the dispersed visibility curves which is only partially probed by our limited dataset.

There is a difference in behaviour between the 8-9 and 9-12 $\mu$m spectral regions that may reflect different opacity regimes, with a decreasing opacity at larger wavelength explaining the higher star and cluster relative fluxes and also an increase of the contribution of the cooler and more extended parts of the envelope.

%We estimate the separation distance from the central star to the clump to be 23 mas, and the P.A. of the clump to be  172$^\circ$. %\textbf{(???errors???)}. 

%However, there is no obvious signature of a clump in the 5 baselines of 2007, covering a much lower spatial frequency range than the 2005 dataset. %\textbf{(???why is the clump not obvious, when we state earlier that the best fit has a clump in it?)} 

\subsubsection{ {\tt DUSTY} Fitting} 
\label{sssec:RYDustyFit}
%For the RY Sgr 2005 dataset the visibilities were fitted for the two shortest baselines only (B$_5$ and B$_6$).  
{\tt DUSTY} can only account for the broad characteristics of the dust shell, and is unable to provide any information on asymmetries  detected. %mostly, in our case, by the longest baselines. % For instance, the change in visibility due to the clump modelled for both the 2005 and 2007 RY Sgr datasets cannot be modelled accurately by {\tt DUSTY}.   
In 2005, the longer baselines (B$_1$, B$_2$, B$_3$, B$_4$) probe the smaller structures (i.e. closer to the star) and show strong evidence of a cluster.  Therefore, these baselines can not be accurately fit with {\tt DUSTY}.    In contrast, the short baselines (B$_5$, B$_6$, B$_7$) should detect the symmetrical Gaussian shell and can theoretically be fit with {\tt DUSTY}.   However, a departure from symmetry is clearly seen when comparing B$_5$ and B$_7$.  Both baselines share the same length, but have different PAs.  Of course, if the circumstellar material was spherically symmetric these measurements would produce the same  visibility, but B$_7$ has a significantly higher visibility than B$_5$.  This implies that an asymmetry also exists at larger scales around the star \citep[see][]{2007A&A...466L...1L}.   %Note also that the two measurements were secured with an interval of one month 
As a result, the reduced $\chi^2$ values for the RY Sgr (2005) {\tt DUSTY} fits (presented in Table \ref{tab:dustyMod}) are calculated from only the  B$_5$ and B$_6$ visibility curves.

%Therefore, thesebaselines can also not be accurately fit with {\tt DUSTY} (see Figure \ref{fig:RY2005Dusty}).   %\textbf{In contrast, B$_7$ deviates from the best geometrical model, hinting towards asymmetry .} %possibly due to a smaller,  second clump at about 30 mas.}  
% \textbf{ This is clear proof that RY Sgr is highly asymmetric  \citet{2007A&A...466L...1L}. } %and implies an extra compact region of dust, which agrees with the suggestion of a second clump by  \citet{2007A&A...466L...1L}.}  
%Additionally, some discrepancies observed between the model and the observed visibility curves probably originate from the complexity of the circumstellar environment which cannot be easily explained by invoking opacity effects.

%Furthermore, {\tt DUSTY} considers the region between the star and the dust shell inner rim to be devoid of any material. This is a strong limitation that may explain why the longest baselines are not well accounted for.  This material may be screening the source or provide some emission.  Moreover, the dust shell inner rim is an artificial simplification of the complex dust forming region in which the flux rises rapidly.

In 2007 only short baseline observations were obtained.  These can be fit relatively well with  {\tt DUSTY}, although B$_4$ and B$_5$ show a hint of asymmetry (the reduced $\chi^2$ values for  {\tt DUSTY} fits are based only on the visibility curves of the 3 shortest baselines).  This agrees with the asymmetry found with the geometrical models.  It is interesting to note that the position of the cluster inferred with good confidence in 2005, and the other suspected in 2007 both lie close to the dust shell inner rim inferred from the  {\tt DUSTY}  modeling.
{\tt DUSTY} places the  dust shell (R$_{{\tt DUSTY}}$)  at $\sim$ 13 mas in 2005 and 17 mas in 2007.  These values are the same, within the uncertainties, as the dust shell inner rim found with the geometrical models ($\sim$ 9 mas in 2005 and 18 mas in 2007).

%However, the geometrical models can only perform a simplified Gaussian model, placing the FWHM somewhere in between the inner and outer radius.  It is not necessarily indicative of the size of the shell or how close the inner radius is.  In this way {\tt DUSTY} gives us more accurate information on the size and location of the shell. Therefore we can say that the geometrical models showing a larger shell in 2007 is due to an opacity effect; in reality, the dust shell is smaller in 2007.  \textbf{ I will investigate how this compares to the light curves.  As Geoff said, it might not be apparent in the light curve because we have no IR monitoring. } 

%\textbf{discuss:  Is that bad data from 12 - 13 $\mu$m? get better spectra}

\subsection{V\,CrA}
\label{ssec:VCrA}

\begin{figure}

  \centering
   % \subfigure[ ] {
  \includegraphics[width=4.58cm]{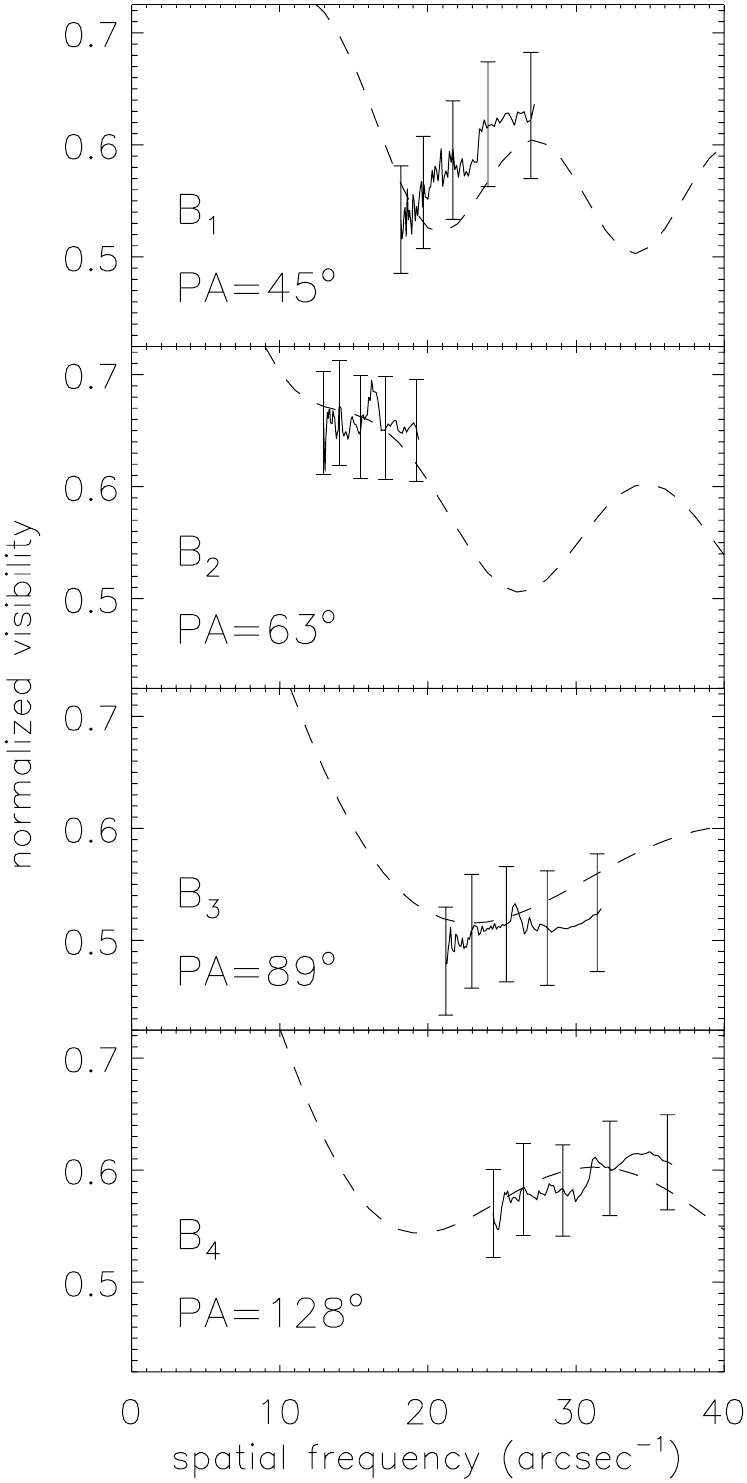}%VCrAfitmono_A.pdf}
%  }
%   \subfigure[ ] {
%  \includegraphics[width=4.cm]{fig5b.pdf}%VCrAvcurves_wl_A.pdf}
%  }
%   \subfigure[ ] {
%  \includegraphics[width=4.cm]{fig5cClus.pdf}%VCrAparams_wl_A.pdf}
%  }

  \caption{ Monochromatic geometrical fits (dashed line) for V CrA using a model consisting of a point source, a Gaussian shell and a cluster. %(a) The fits using the the monochromatic models are shown. % (b) Fits using the the chromatic models are shown.  (c) Corresponding chromatic quantities derived from the best-fitting model. Each panel is labelled by the baseline and PA referenced in Table \ref{tab:VLTIlog} and Figure \ref{fig:uvplane}. The observed visibility curves (solid lines) have indicative error bars marked.  The best fit to the observed visibilities is marked by the dashed line. 
    \label{fig:VCrAGeo}}
    % \label{fig_vis1*}
   \end{figure}

 \begin{figure*}

\begin{center}
 % \centering
    \subfigure[ ] {
  \includegraphics[width=5.95cm]{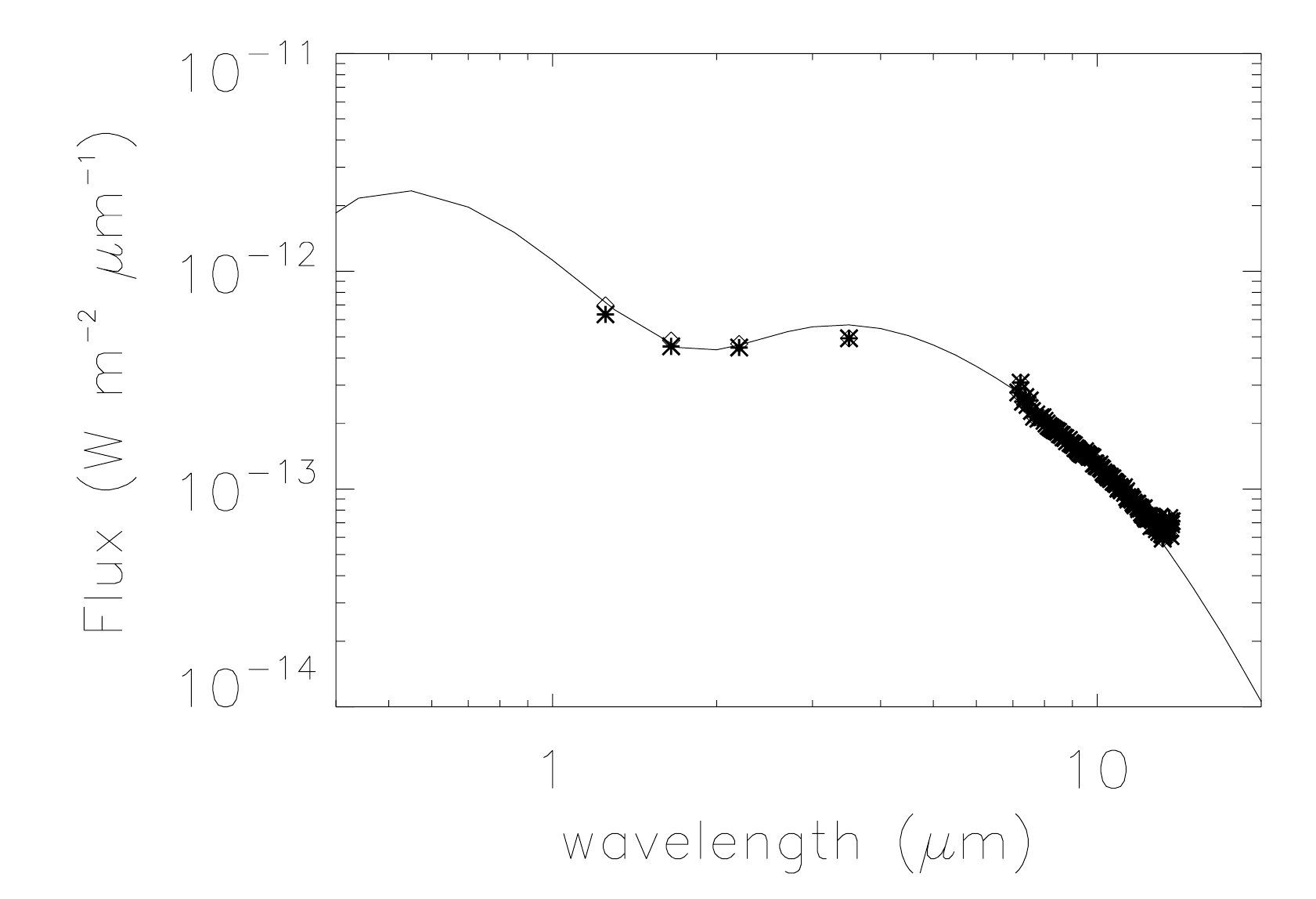}%VCrAfitmono_A.pdf}
  }
   \subfigure[ ] {
  \includegraphics[width=5.5cm]{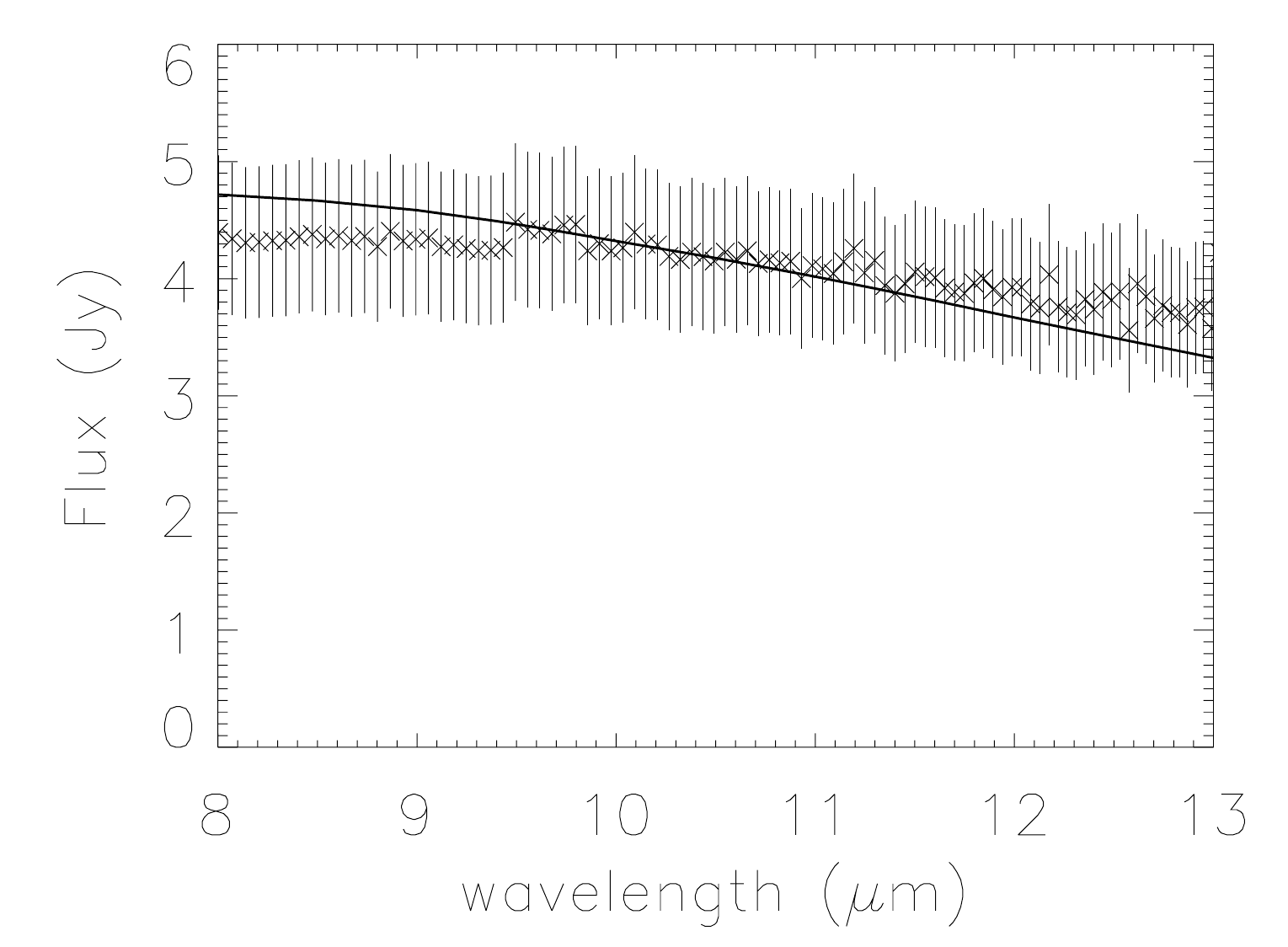}%VCrAvcurves_wl_A.pdf}
  }
   \subfigure[ ] {
  \includegraphics[width=5.5cm]{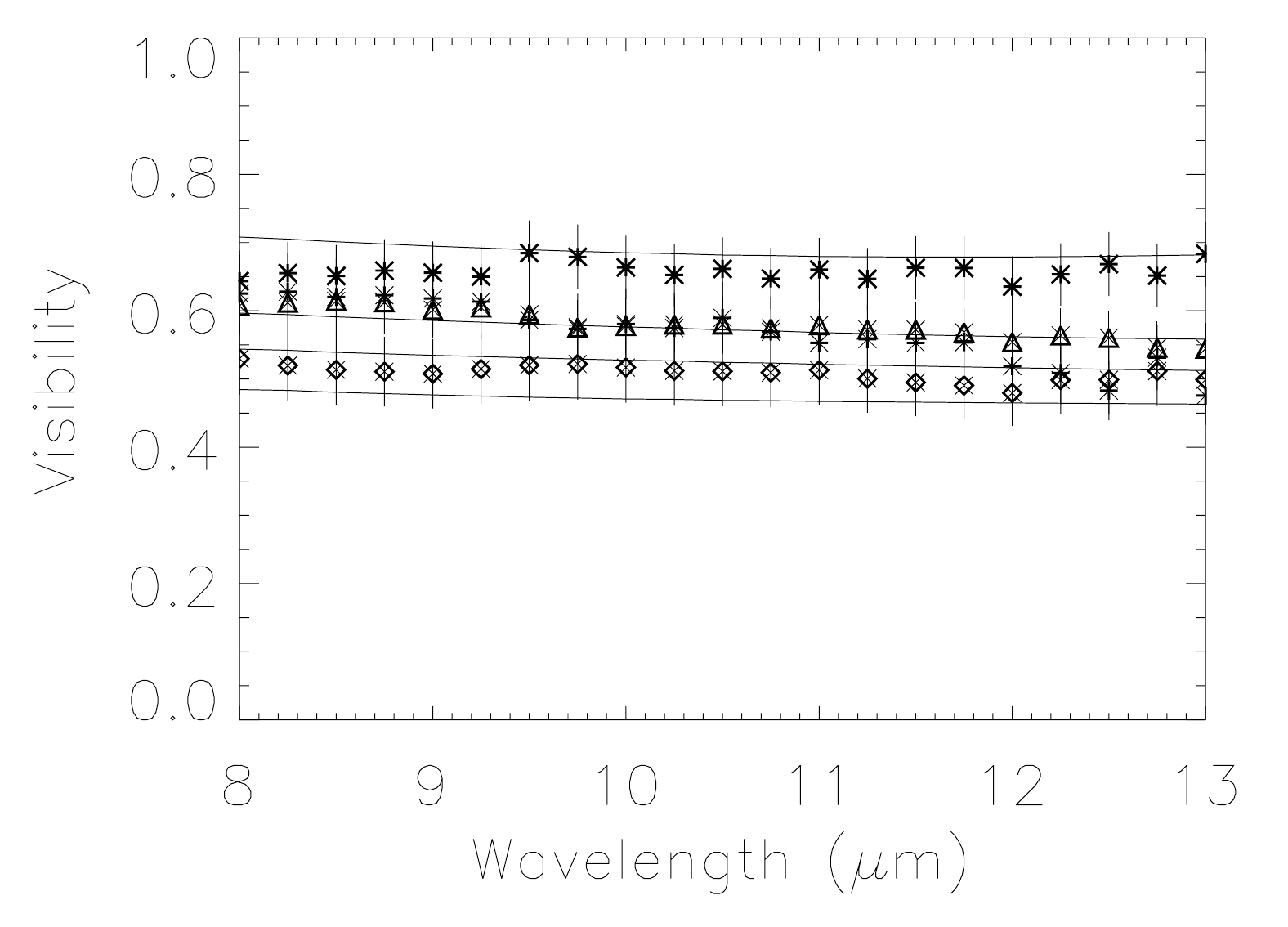}%VCrAparams_wl_A.pdf}
}

     \caption{Best {\tt DUSTY} fits for V CrA.  (a) The modelled SED (solid line) is shown along with the observed spectra (symbols). (b) Only the MIDI spectrum  (crosses) is shown with the model (solid line). (c) The 4 baseline visibility curves (symbols) are shown with the visibility fits (solid line). The visibility curves are shown from the shortest baseline (top) to the longest baseline (bottom). Indicative error bars are marked in each figure.   \label{fig:VCrADusty}}
  \end{center}	
   \end{figure*}
   
\subsubsection{Geometrical Fitting}
\label{sssec:VCrAGeoFit}
%\textbf{ from Olivier: The V CrA observations are limited to 4 baselines only, and it is not possible in this case to find a unique solution, as a strong degeneracy between the models is observed. A monochromatic model with a central source and a Gaussian disk provides already a good fit of the visibilities. The Gaussian has a FWHM of 46mas, and the star contributes to 56\% of the total flux. Including a clump helps to find better $\chi$$^2$ for this limited dataset, but assuming a 2D Gaussian ellipsoid improve the fit similarly. In any case, the flux from the clump is not larger than 5-6\% of the total flux. Chromatic fitting are not satisfactory as well. }

%\textbf{ -- but I don't see this to be the case.  the PS+G visibility fits look very poor compared to the better fits of PS+G+clump)}

Similar to RY Sgr, the geometrical models for V CrA reveal that the visibility curve of this source is best fit by a combination of a central star, a shell, and a dust cluster (see Figure \ref{fig:VCrAGeo}).  %The $\chi^2$ for this model (0.21) is slightly better than a fit with only a central star and Gaussian shell ($\chi^2$=0.81).   
We find the separation of the cluster and the star to be 84 mas (437 AU, using a distance of 5200 pc)  and the PA of the cluster to be 16$^\circ$.  The radius of the Gaussian shell is 22 mas.  This places the cluster about 4 times farther away than the shell's radius.  In addition, the dust cluster has a flux of only 5\% of the total system flux.  However, all the geometrical models  for V CrA provide a similar reduced $\chi^2$.  Therefore the inclusion of a cluster is not clearly justified by the data.
%Therefore, we cannot definitely conclude whether or not there are asymmetries found in the circumstellar material of V CrA.

%The wavelength-dependent analysis \textbf{(Figure \ref{fig:VCrAGeo}c)} does not provide any evidence of change of the parameters with wavelength, except perhaps in a narrow spectral region close to 8 $\mu$m. The best geometrical fits for V CrA are found in Figure \ref{fig:VCrAGeo}.

\subsubsection{ {\tt DUSTY}  Fitting}
\label{sssec:VCrADustyFit}
Good {\tt DUSTY} fits for V CrA were produced for the SED and four visibility curves.  The consistency between the visibility curves and the SED suggest that the large estimated distance to V CrA (5200 pc) is appropriate.  The consistency may also suggest that a spherical shell surrounding the star may be an appropriate model for V CrA.  However, because V CrA is so far away, it is not the best object to detect asymmetries in the circumstellar environment.   The general shape of the target can be seen, but the inner circumstellar regions where asymmetries are more likely to be found cannot be resolved.      

{\tt DUSTY} places  the dust shell (R$_{{\tt DUSTY}}$) at 5 mas.  Unlike RY Sgr,  this is significantly smaller than the inner rim predicted by the geometrical models for V CrA ($\sim$ 20 mas).
 The {\tt DUSTY}  fits to the V CrA data are found in Figure \ref{fig:VCrADusty}.

%\textbf{Add more detail -- What is different about V CrA?  Why doesn't it have any dust features like RY Sgr or V854 Cen?} 
%
% \begin{figure}
%  \centering
%  \includegraphics[width=9cm]{VCrALC.pdf}
%  %\includegraphics[width=7.8cm]{Fig6_LeaoPart1.ps}
%  \caption{ V CrA light curve from AAVSO.  Allowing for errors in the dust's velocity the clump observed with the VLTI in June 2007 could have been produced during the same period of activity that caused the declines in V CrA's light curve approximately 8-12 years prior to observations. }
%    % \label{fig_vis1*
%   \end{figure}

\subsection{V854\,Cen}
\label{ssec:V854Cen}

\subsubsection{Geometrical Fitting}
\label{sssec:V854CenGeoFit}

%\begin{itemize}
%\item{I have included a rough fit for V854 Cen (Figure 11).  I am still figuring out the plotting.  This is the same scale as Izan's other plots. }

For V854Cen only two baselines were obtained.  As a result the geometrical fits are highly degenerate.  In addition, because the baselines obtained are short, they only probe the larger scale structure, or larger shell portion of V854 Cen.  Therefore, we used only the two most basic models, a Gaussian shell, and a Gaussian shell plus cluster to fit the 2 baselines (see Table \ref{tab:geoMod}).  %In this case the minimum $\chi^2$ for a central star and a shell (0.58) is slightly better than that of the central star, shell, and clump ($\chi^2$=0.8).  
However, the visibility curves for the two fits are too similar to determine if there is any asymmetry.  We therefore only use the Gaussian shell fit for our analysis (see Figure \ref{fig:V854CenGeo}). This places the FWHM of the Gaussian shell at 24 mas with a flux comprising of 84\% of the total system.  %The wavelength-dependent analysis \textbf{(Figure \ref{fig:V854CenGeo}c)} does not provide any evidence of change of the parameters with wavelength.  The geometrical fits for V854 Cen are found in Figure \ref{fig:V854CenGeo}

%\item{I haven't compared this to the  {\tt DUSTY}  results yet. } 
%\end{itemize}

\subsubsection{ {\tt DUSTY}  Fitting}
\label{sssec:V854CenDustyFit}

Good {\tt DUSTY} fits for V854 Cen are obtained for the 2 baselines, and asymmetries are not implied from the fits obtained.  %, with a $\chi^2$ of 3.0 for the visibility fit 
But at  $\sim$8 $\mu$m and $\sim$11.5 $\mu$m an emission feature is seen in both the SED and the visibilities that causes the model to deviate from the observations. These features are discussed in more detail in \S\ref{sec:discuss}.   

{\tt DUSTY} places the dust shell at 11 mas (R$_{{\tt DUSTY}}$).  This is similar to the inner rim predicted by the geometrical models for V854 Cen (12 mas).
 The {\tt DUSTY}  fits to the V854 Cen data are found in Figure \ref{fig:V854CenDusty}.

%Using the Spitzer/IRS spectrum. From $\sim$7-9.5 $\mu$m we clearly see evidence of an extended HAC emission feature and we can also see a small feature at $\sim$11 $\mu$m (discussed in Lambert 2001).  The best  {\tt DUSTY}  fits for V854 Cen are found in Figure 9.
%The dip seen in the visibility curve at 8 $\mu$m  is caused by the HAC blocking part of the core visibility source.  Visibility curves are heavily sensitive to changes in the core source (because it is a compact object); since the visibility of the HAC is 0 (as it is over resolved) and it is blocking part of the core, the total visibility has decreased below the model at 8 $\mu$m by 19$\%$.  Likewise, the amount of extra flux that the extended HAC contributes to the SED is 20$\%$ at 8 $\mu$m.  Therefore we are seeing a drop in the visibility curve which is proportional (within given errors) to the increase of flux in the SED.      

%
\begin{figure}

  \centering
  %\includegraphics[width=7.8cm]{Fig6_LeaoPart1.ps}
  % \subfigure[ ] {
  \includegraphics[width=4.65cm]{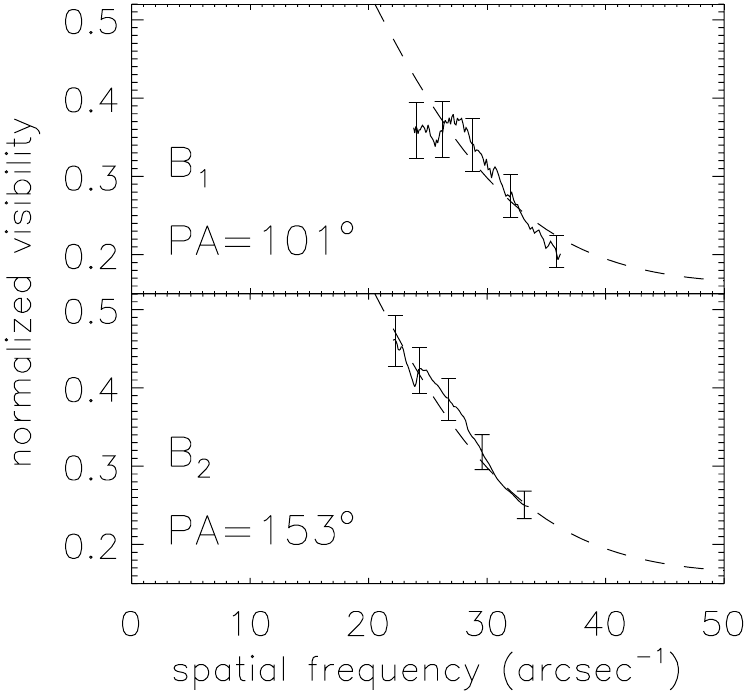}%VCrAfitmono_A.pdf}
  %}
  % \subfigure[ ] {
 % \includegraphics[width=4.cm]{fig_v854_b.pdf}%VCrAvcurves_wl_A.pdf}
 % }
 %  \subfigure[ ] {
  %\includegraphics[width=3.8cm]{fig_v854_c.pdf}%VCrAparams_wl_A.pdf}
  %}

  \caption{ Monochromatic geometrical (dashed line) fits for V854 Cen using a model consisting of a point source and a Gaussian shell. %(a) The fits using the the monochromatic models are shown.  (b) Fits using the chromatic models are shown.  (c) \textbf{Corresponding chromatic quantities derived from the best-fitting model.} Each panel is labelled by the baseline and PA referenced in Table \ref{tab:VLTIlog} and Figure \ref{fig:uvplane}. The observed visibility curves (solid lines) have indicative error bars marked.  The best fit to the observed visibilities is marked by the dashed line. 
    \label{fig:V854CenGeo}}
    % \label{fig_vis1*}
    
   \end{figure}

\begin{figure*}

\begin{center}
  \subfigure[ ] {
  \includegraphics[width=5.9cm]{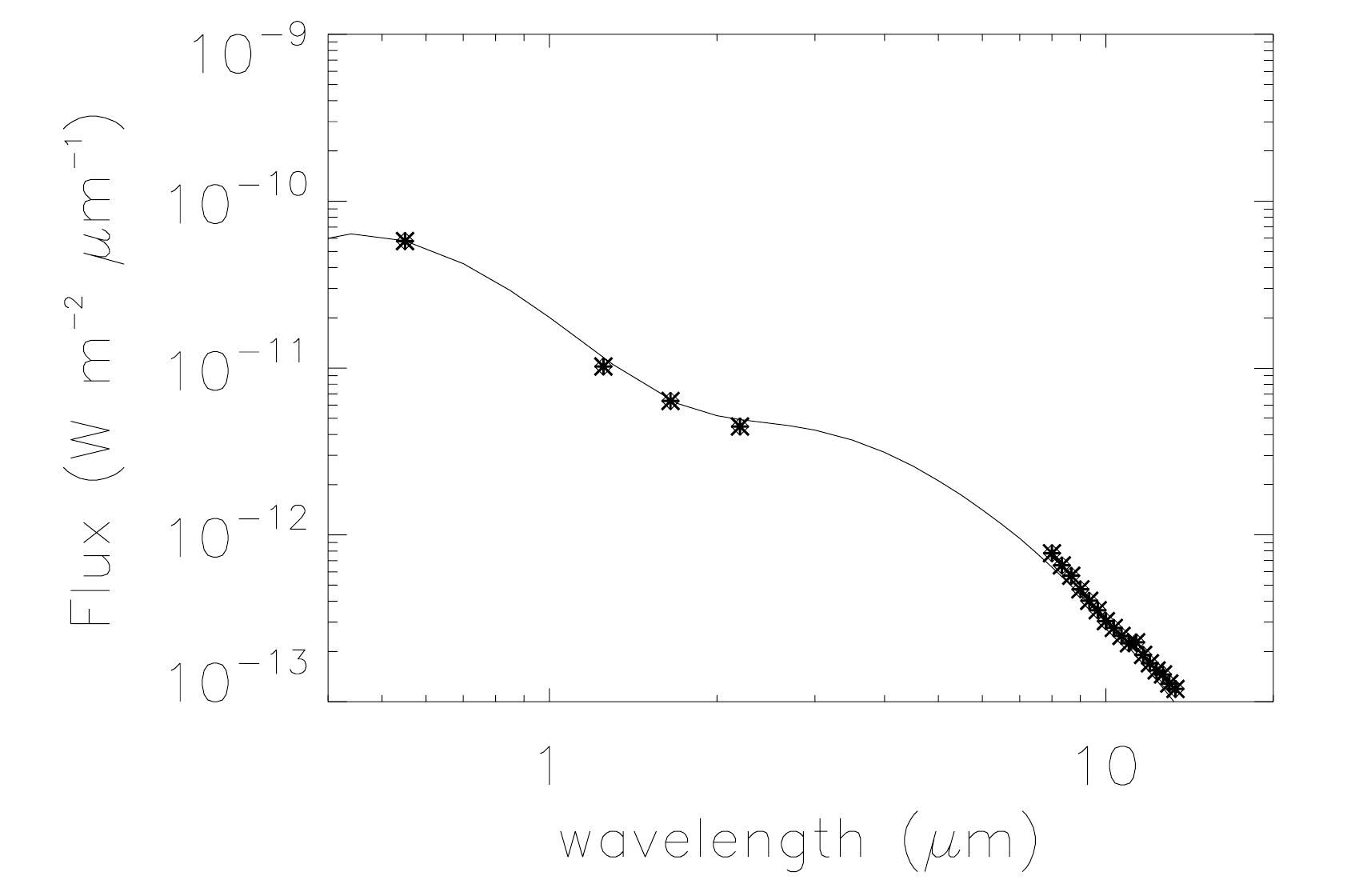}%VCrAfitmono_A.pdf}
  }
   \subfigure[ ] {
  \includegraphics[width=5.5cm]{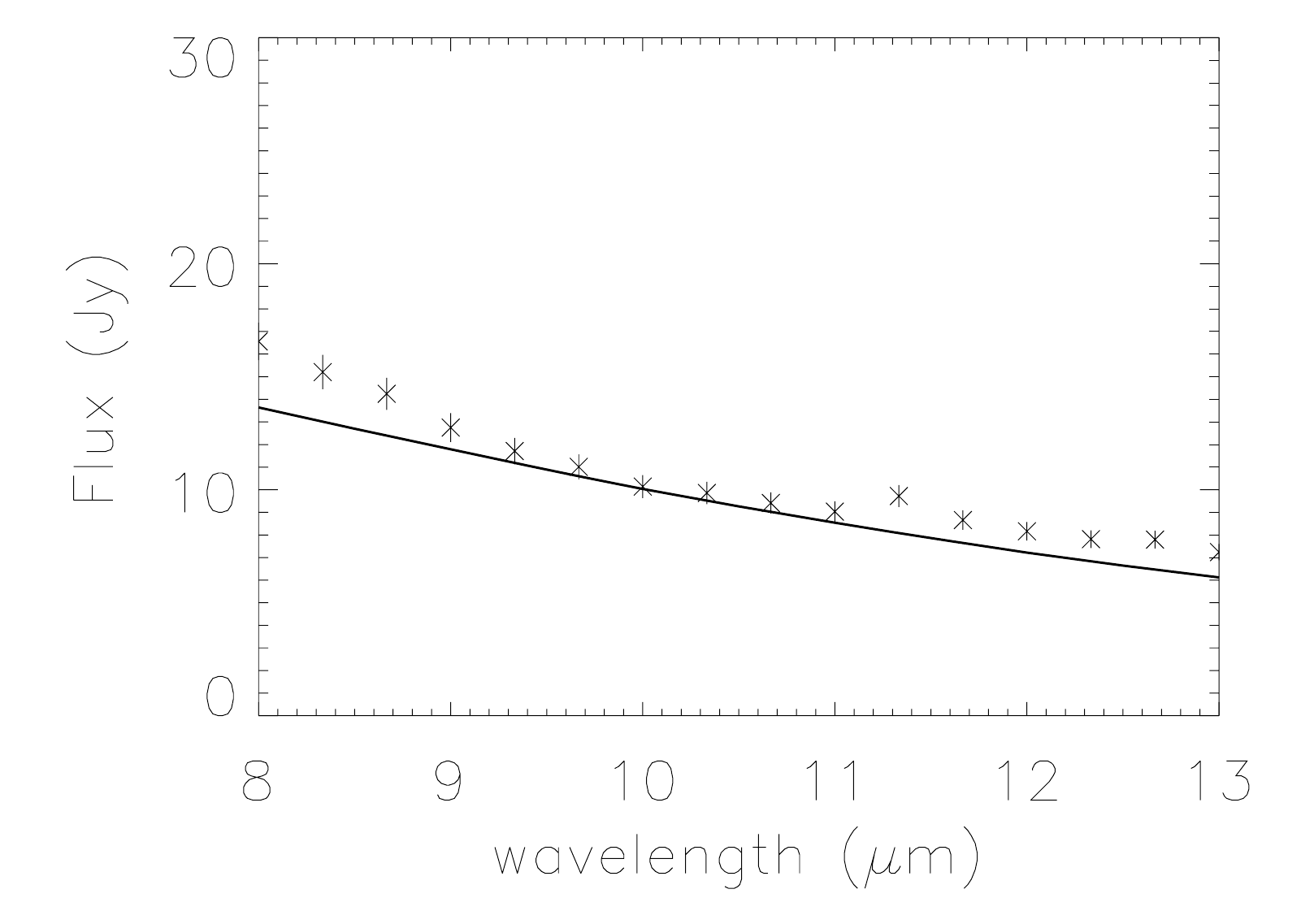}%VCrAvcurves_wl_A.pdf}
  }
   \subfigure[ ] {
  \includegraphics[width=5.5cm]{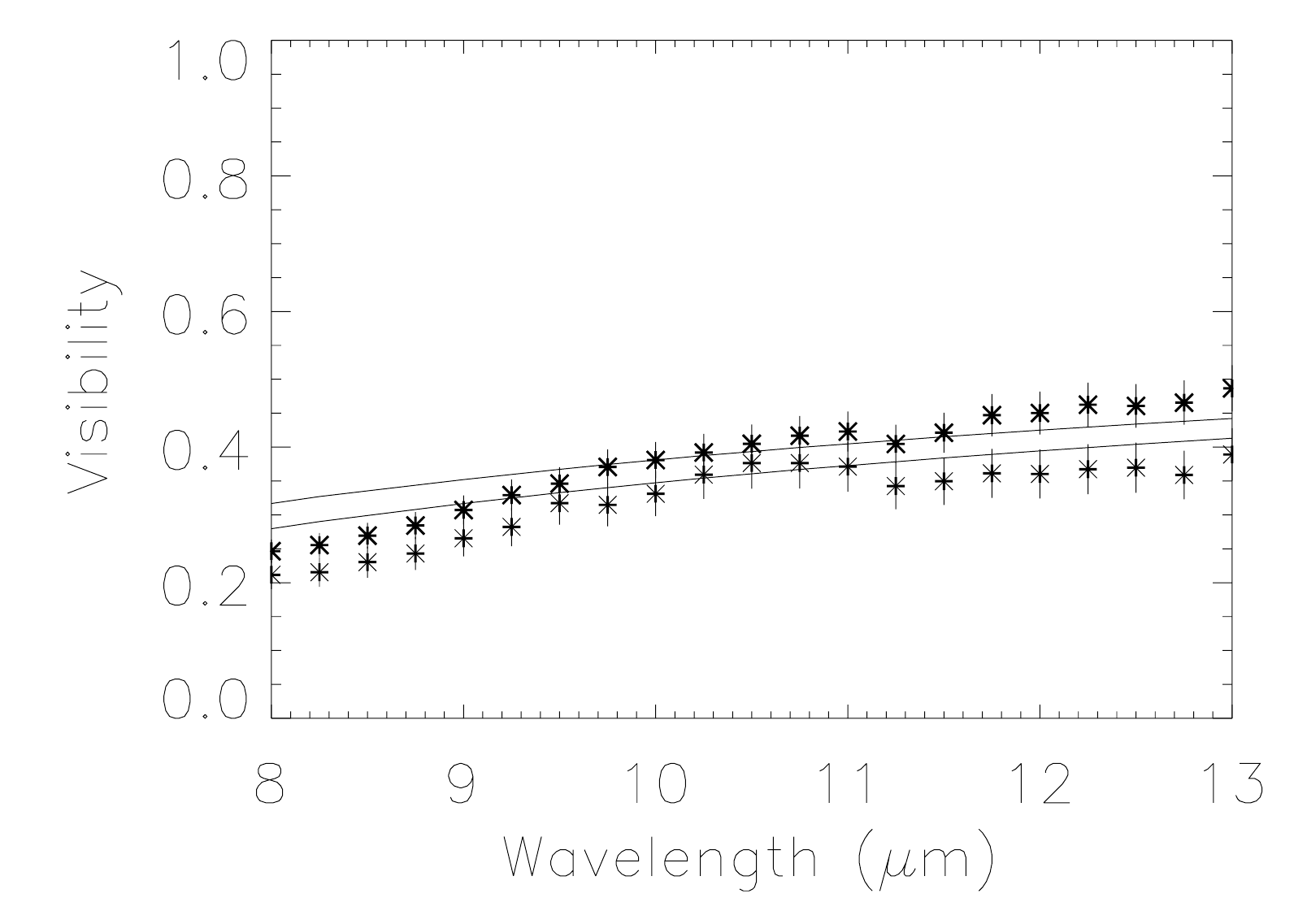}%VCrAparams_wl_A.pdf}
  }
     \caption{Best {\tt DUSTY} fits for V854 Cen.  (a) The modelled SED (solid line) is shown along with the observed spectra. (b) Only  the MIDI spectrum (crosses) is shown with the model (solid line). (c) The 2 baseline visibility curves (symbols) are shown with the visibility fits (solid line). Indicative error bars are marked in each figure.     At $\sim$ 8  and 11.5 $\mu$m we see evidence of two extended emission features. 
          \label{fig:V854CenDusty}}
  \end{center}	
   \end{figure*}

\section{Discussion}
\label{sec:discuss}

%This section determines how our observations and modeling results fit within current knowledge of RCB stars.  Specifically we seek to find what the results reveal about dust production activity for these stars and if our results suggest a preferential plane of dust ejection.

\subsection{Dust Formation Frequency}
\label{ssec:Frequency}

\begin{table*}
\begin{caption}
{Time for dust to reach shell and mass of shell
\label{tab:timemass}}

\end{caption}

\begin{tabular}{llccccc}\hline 
Star & $R_{{\tt DUSTY}}$ & t$^a$ & Pulsational Period & $M_{shell}$ & $n_{clumps}$  \\
&  (mas), (\rstar) & (months) & (days) & (\msol)         \\ 
\hline
RYSgr 2005 & 13, 79 & 5& 38$^b$ & 3.7 x 10$^{-7}$ & 37  \\
%\hline
RYSgr 2007 & 17, 108& 7 & 38$^b$ & 3.7 x 10$^{-7}$ & 37 \\
%\hline
V CrA & 5, 100 & 5& 57$^c$ &6.5 x 10$^{-7}$ & 65   \\ 
%\hline
V854 Cen & 11, 71 & 5 & 43$^b$  &  1.4 x 10$^{-7}$ & 14    \\ 
\hline

\end{tabular}

{\tiny $^a$ Time to reach $R_{{\tt DUSTY}}$, see \S\ref{ssec:DustyFit} 

$^b$From  \citet{2007MNRAS.375..301C}

$^c$From  \citet{1990MNRAS.247...91L}
}
\end{table*}

Our observations, in combination with optical light curves, may shed light on the frequency and location of dust formation in RCB stars.  Current knowledge suggests that  dust forms in clumps at discrete locations around RCB stars and a visible light decline will be observed if a clump is ejected in the line of sight. Therefore, declines in the lightcurve will not necessarily  correspond to structure observed by the VLTI. Furthermore, if a clump is ejected towards us, causing a decline in the visual light curve, it is unlikely to be detected by the VLTI. Conversely, a clump that is easily detected by the VLTI is likely to be one that was not ejected towards us and therefore did not produce a lightcurve decline. On the other hand, RCB stars are known to undergo periods of inactivity, where the star is at maximum light sometimes for years, and periods of strong activity when the star is constantly declining and recovering \citep{1996PASP..108..225C}. Because of this, we presume that the clusters we resolve were likely produced during a period of high dust production activity that also resulted in light declines.%Because of this, we may presume that the clumps we observe and that were ejected perpendicular to the line of sight, might have been ejected in a dust production episode which also caused a decline in the lightcurve.

Using the distance to each of our RCB stars and their stellar radii, as well as an average dust outflow velocity of 300 km/s, we estimate the time it takes a dust clump to reach the observed dusty shell (R$_{{\tt DUSTY}}$, see \S\ref{ssec:DustyFit}) after initial ejection, assuming the dust forms at 2 \rstar \citep{1992ApJ...397..652C,2003ApJ...595..412C}.
As discussed in \S\ref{ssec:GeoFit}, the geometrical cluster represents a break from symmetry in the distribution of dust clumps of the shell, but does not necessarily indicate the position of a particular clump. Therefore, we use the simplistic R$_{{\tt DUSTY}}$ as the distance for an individual clump to travel for our analysis.   We also assume that all the dust leaves the star in the plane of the sky; therefore this estimated time-scale is actually a lower limit.  Table \ref{tab:timemass} lists R$_{{\tt DUSTY}}$, and a lower limit of the time it takes the dust to reach R$_{{\tt DUSTY}}$ for each of our RCB stars.

V-band light curves of our RCB stars are plotted in Figure \ref{fig:LC} using data from ASAS-3 \citep{2002AcA....52..397P} and the AAVSO.  The 2005 and 2007 VLTI observations of RY Sgr were obtained on JD 2453550 and JD 2454275. Dust was observed at 13 mas and 17 mas (24 AU and 31 AU, see Table \ref{tab:dustyMod}) in 2005 and 2007, respectively.  It will take a minimum of 5 and 7 months for the dust to reach these distances.  Five months prior to the 2005 VLTI observation, when the dust was expected to be ejected, there is a decline in the light curve (JD 2453330).  However, seven months prior to observation of RY Sgr in 2007 (JD 2454065), when the dust was expected to be ejected in 2007, there is not a visual decline. %{\bf but there is a stretch of time with no data, is this so?}  
There are no declines during the period of JD 2453550 - 2454500. %It is difficult to conclude whether the dust we observed is connected to activity that produced the light curve declines. 

Dust was observed at 5 mas (1 AU) around V CrA in 2007.  It would take at least 5 months from ejection to reach this distance.   Looking at the light curve five months prior to our observation (JD 2454130), V CrA is just beginning a visual decline after a long period of inactivity (at least 7 years). While the light curve shows relative inactivity prior to this time (especially compared to V854 Cen) V CrA still has significant IR emission.  This could be an indication that dust around V CrA is actively forming in a plane that is not in our line of sight. 

Dust was observed at 11 mas (5 AU) around V854 Cen in 2007.  It would take at least 5 months from ejection to reach this distance. Five months prior to our observation (JD 2454130), V854 Cen is in the middle of an extremely active dust production period, such that the observed dust cluster is likely to be related to those that are obscuring our line of sight. 

Without additional epochs of VLTI observations of these RCB stars, covering a significant time period, it is difficult to reconstruct the dust production frequency and history of these objects. But it is clear from our analysis that observations such as these, taken over time, combined with IR and visible brightness variations, could allow us to determine the time and location of dust production in these objects.

\begin{figure*}
\centering
\includegraphics[width=14cm]{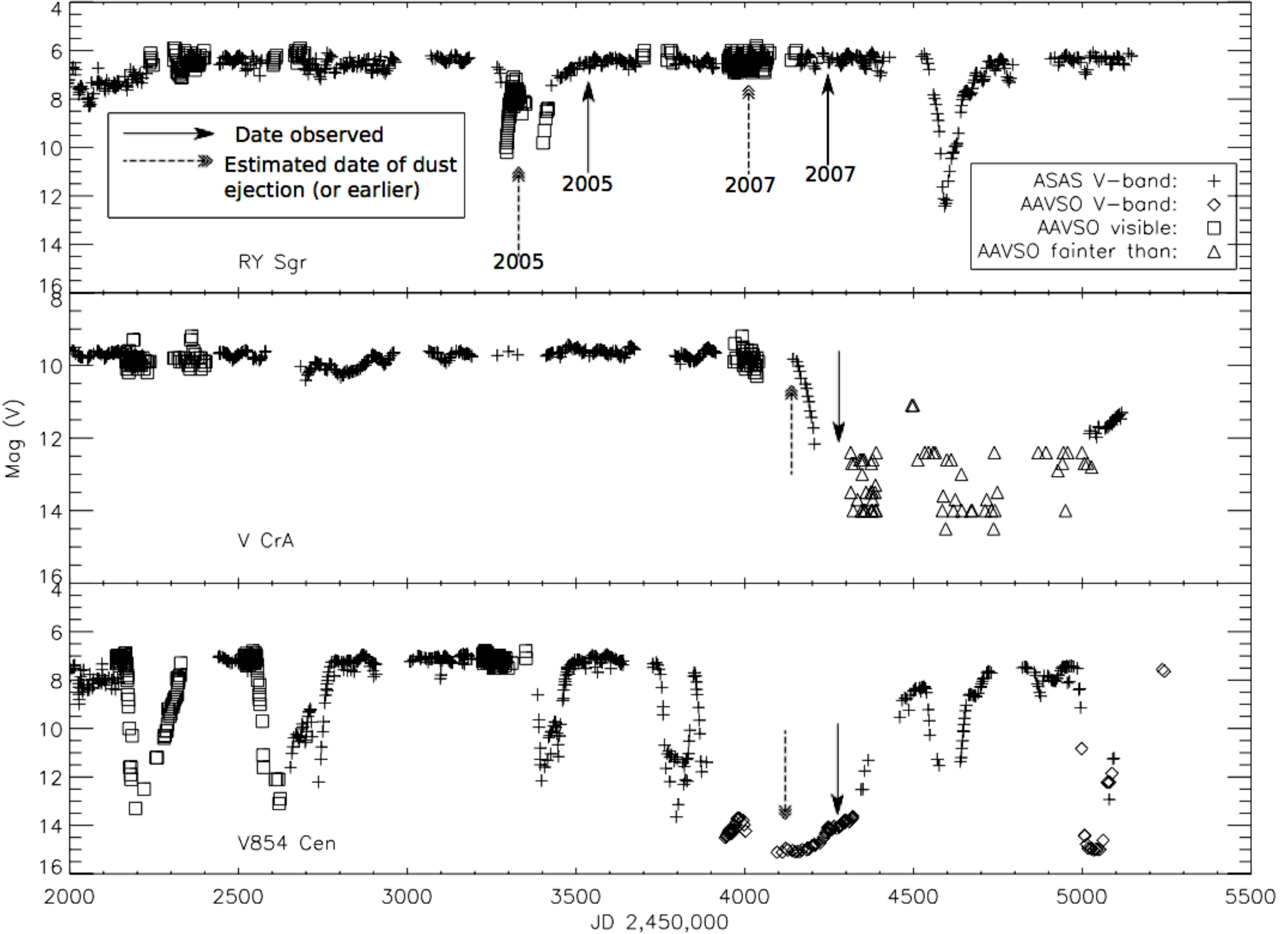}
\caption{ Light Curves for RY Sgr (upper), V CrA (middle), and V854 Cen (lower), from April 2001 to Jan 2010 are shown.  Our VLTI observations took place in June 2005 (RY~Sgr; JD 2453550) and June 2007 (RY~Sgr, V~CrA and V854~Cen; JD 2454275). Estimated dates when the observed dust was ejected are marked with arrows.  The intervals between ejection and observation should be considered lower limits.  
\label{fig:LC} }
\end{figure*}

\subsection{Dust Mass}
\label{ssec:Mass}

%According to studies of RCB visual light curves (e.g. \citealt{1992ApJ...397..652C}), after the initial decline there is often a fast rise of $\sim$ 0.13 mag per day.  For an initial 8 magnitude decline it will take $\sim$ 60 days for the star to return to maximum light.  With  a typical dust flow velocity of 300 km/s, dust will become optically thin at $\sim$ 30 \rstar (\rstar = 85~R$_\odot$)~ and will no longer cause light extinction \citep{1992ApJ...397..652C}. These considerations are consistent with our best {\tt DUSTY} model's dust shell radii (R$_{{\tt DUSTY}}$, see Table \ref{tab:dustyMod})  of $\sim$ 90 \rstar.  %This also supports the assumption that we can use optically thin dust to calculate the shell dust mass.

MOCASSIN, a 3D Monte Carlo code \citep{2005MNRAS.362.1038E}, was used to model the radiative transfer in the shells surrounding our three RCB stars, resulting in fits to the three MIDI SEDs. (Note that the RY Sgr MIDI 2005 and 2007 data were identical within 5\%, so only the 2007 data is used in this paper.)  The same input parameters were used as in the {\tt DUSTY} fits listed in Table \ref{tab:dustyMod}. Using MOCASSIN we find  that the dust shells have an average mass of  $\sim$ 4 $\times$ 10$^{-7}$ M$_{\sun}$.   \citet{1992ApJ...397..652C} estimated that a single ejected clump has a mass of $\sim$ 10$^{-8}$ M$_\odot$.  Therefore, using a simplified calculation, we find that for each shell to grow in mass to 4 $\times$ 10$^{-7}$ M$_{\sun}$ it has to contain approximately 40 individual clumps.  The time it takes to accumulate this amount of mass may be estimated by using the pulsational period of the RCB stars. 

The onset of dust formation has been found to be correlated with the pulsational phase in five RCB stars, including RY Sgr and V854 Cen \citep{2007MNRAS.375..301C,1999AJ....117.3007L,1977IBVS.1277....1P}. RCB stars pulsation amplitudes are at most a few tenths of a magnitude and their periods are 40-100 d \citep{1997MNRAS.285..266L,1990MNRAS.247...91L}.  If one clump is ejected every pulsation period (or roughly every 50 days) it will take $\sim$ 5 years to build a shell with the typical observed masses.  If the dust expands at 300 km/s, a clump will be at $\sim$  800 \rstar~  from the star after five years, which is  well within the highly unconstrained, outer shell radius estimated by {\tt DUSTY}, $\sim$ 3000 \rstar (assuming, as before, R$_{{\tt DUSTY}}$ = 10 mas, and a distance of 2000 pc). %\textbf{? Does this last statement seem okay?, 800 seems rather small compared to 3000}.  
Table \ref{tab:timemass} contains specific calculations for our RCB stars, including the pulsation period for each star, the mass of its shell, and the number of clumps within the shell.

\subsection{Dust Geometry}
\label{ssec:DustGeo}

Looking at our {\tt DUSTY} results in connection with the light curves,  the time it takes for ejected dust to travel to R$_{{\tt DUSTY}}$ (see Table  \ref{tab:timemass}), and the dust shell mass,  we find that, overall, our observations are consistent with dust forming in clumps ejected randomly around the star so that over time they may create a spherically symmetric distribution of dust.  But  we also must look to our geometrical models to determine if there is an overall preferential dust ejection axis or plane.

The geometrical fits (see \S\ref{sec:Results} and Table \ref{tab:geoMod}) show evidence for asymmetry in the dust geometry around at least one of our RCB stars.   Looking at the visibility curves (Figure \ref{fig:RY2007Geo}a, \ref{fig:VCrAGeo}, \ref{fig:V854CenGeo}), asymmetric structure is apparent in RY Sgr, may also exist in V CrA,  and is possible for V854 Cen but is not constrained at present (see \S\ref{sec:modeling}). Our geometrical models provide a position angle (PA) of the asymmetric distribution of dust (or cluster) and by observing these objects over time, we can learn  whether there is a preferential dust ejection axis or plane.  Although the geometrical models only give the simplistic version of one large ``cluster", this represents an overall asymmetry found in the production of dust.

Only RY Sgr has more than one epoch of data and asymmetries are detected in both 2005 and 2007.  The RY Sgr 2005 model has an asymmetry placed at a PA  of 79$^\circ$ and in 2007 the PA of the asymmetry is at 175$^\circ$, approximately 90$^\circ$ from the 2005 epoch.  It is important to note that although the geometrical model, comprised of a star, a Gaussian shell and  a cluster, gives us the lowest reduced $\chi^2$ value, the same PAs are also found if an elliptical uniform disk is used implying there is indeed some sort of asymmetry in this direction.    %According to the light curves, we already know that dust is not ejected exclusively on one plane. 
However, the determination of whether there is an overall preferential plane will have to wait until a time series of observations are obtained that measure the location of several ``clusters" over time.

According to the {\tt DUSTY} models, RY Sgr, V CrA and V854 Cen all have dust shells with approximately the same R$_{{\tt DUSTY}}$ despite other differences.  This distance  may be a common characteristic between RCB stars or it may be a bias of the simplistic {\tt DUSTY}  approach to modelling these complex stars.  An insight to this question will require more data and investigation.

\subsection {Large Scale Dust Distribution and Composition}
\label{ssec:Comp}

MIDI spectra were obtained subsequently to the interferometric data.  These spectra correspond to the  flux from the central 250 mas of the sources.    In addition to the MIDI spectra, ISO/SWS spectra exist for RY Sgr and V854 Cen and Spitzer/IRS data exist for V CrA and V854 Cen.  ISO/SWS records flux through a slit with an aperture of 14\arcsec\ x 20\arcsec\ \citep{2003ApJS..147..379S} and Spitzer/IRS  SL spectra  were taken through a 3\arcsec\  wide slit with very long length \citep{2004ApJS..154...18H}.     By comparing these spectra, we can determine where the majority of the flux is coming from.  For RY Sgr  the 2007 MIDI spectra  is $\sim$ 20\% fainter than the 1997 ISO/SWS spectra  (the RY Sgr MIDI 2005 and 2007 data were identical within 5\% so only the 2007 data is used in this paper).   For V CrA the 2007 MIDI  spectra is $\sim$ 30\% brighter than the 2005 Spitzer/IRS  spectra.   These changes are within the expected error bars for the respective instruments (e.g. MIDI spectra can change by 10 - 15\% within a few minutes depending on fluctuations in the atmosphere \citep{2007NewAR..51..666C}.  In addition, the flux from  the stars will have varied by an unknown amount between the ISO, Spitzer, and MIDI epochs. Taking this into consideration we determine that the majority of the flux is coming from the inner 250 mas of the source for both RY Sgr and  V CrA.  On the other hand, for V854 Cen the  the 2008 Spitzer/IRS  spectra is $\sim$ 55\% brighter than the 2007 MIDI  spectra and the 1996 ISO/SWS spectra is  $\sim$ 165\% brighter than the 2007 MIDI  spectra).  These higher percentages cannot be accounted for by uncertainties  or normal variability.    This implies that there is a significant amount of circumstellar material outside of the region observed with MIDI, possibly in another outer shell. V854 Cen is  known to have a 5\arcsec\ circumstellar nebula at UV and mid-IR wavelengths \citep{2001ApJ...560..986C, LagadecPerCom2010}. %Recent, mid-IR images of V854 Cen show a spherically symmetric shell outside the MIDI aperature (Lagadec, personal communication).  
The comparisons of all spectra obtained for the RCB stars are shown in Figure \ref{fig:Spectra}.

\begin{figure}
\centering
\includegraphics[width=8.8cm]{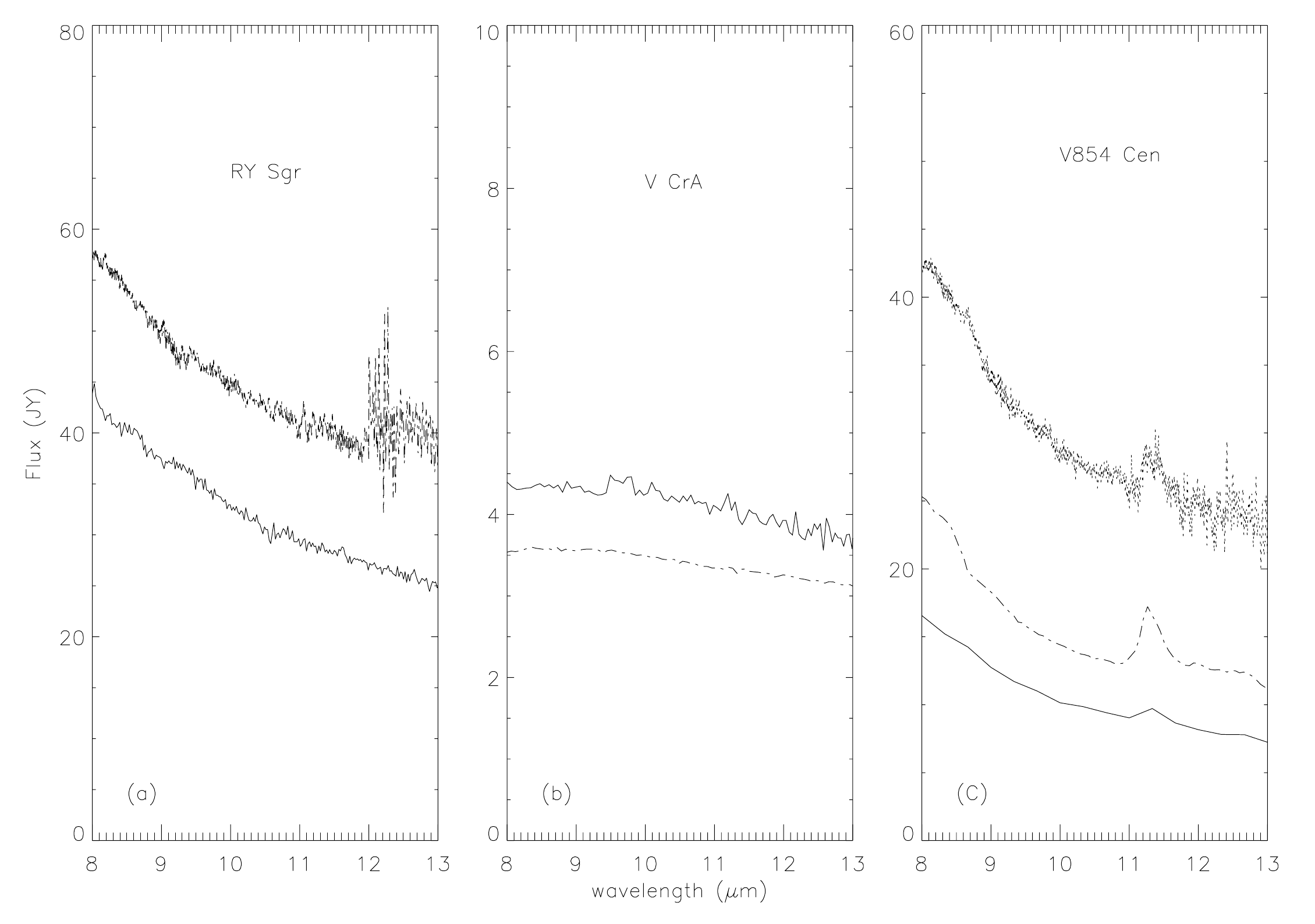}
\caption{ (a) Spectra of RY Sgr: MIDI, lower curve, ISO, upper curve. (b) Spectra of V CrA: Spitzer, lower curve, MIDI, upper curve.  (c) Spectra of V854 Cen: MIDI, lower curve, Spitzer, middle curve, ISO, upper curve. \label{fig:Spectra}  }
\end{figure}

Generally RCB stars, including R CrB and RY Sgr, show no features in the mid-IR \citep{2001ApJ...555..925L, 2005ApJ...631L.147K}.  The exception is V854 Cen which shows polycyclic aromatic hydrocarbons (PAH)-like emission features \citep{2001ApJ...555..925L, 2010ApJ...714..144G}.  Our MIDI spectra for RY Sgr and V  CrA appear to be featureless, but the spectrum of V854 Cen show emission features from $\sim$ 8 $\mu$m and at $\sim$11.5 $\mu$m. These features may be due to hydrogenated amorphous carbon (HAC), C$_{60}$, or PAHs \citep{Evansinprep2011, 2010ApJ...714..144G, 1995AJ....109.2096C}.  RCB stars are known to have variable emission features and C$_{60}$ may be present only for a short time after initial decline when carbon gas is condensing into dust \citep{1993ASPC...45..115W,1992A&A...265..216G}.
These emission features have an impact on V854 Cen's visibility curves (Figure \ref{fig:V854CenGeo}) which exhibit a drop at the features' wavelengths. The dust responsible for the features is over-resolved, i.e., more extended. In other words, dust associated with the emission feature is blocking light from the central unresolved source and the total visibility is lower than the model at 8 $\mu$m by 21$\%$.  This drop is proportional to the amount of extra flux that the feature contributes to the SED (23$\%$) at 8 $\mu$m.  A smaller drop in visibility and excess in the SED is also noticed at 11.5 $\mu$m.   

Our observations with the VLTI have shed light on both the large and small scale structure and composition of RCB stars.  The large extended shell and emission features found for V854 Cen are atypical for RCB stars and need to be investigated further.
In addition, RY Sgr's visibility curves, in both 2005 and 2007, may reveal a slight drop at 8-9 $\mu$m, but is not as evident as the drop for V854 Cen.  This hints at the ability of high-angular resolution observations to detect chemical structure that cannot be revealed by spectrometry. 

\subsection{Dust Grain Size}
\label{ssec:Grain}

Standard MRN  dust distribution \citep{1977ApJ...217..425M}  uses $a_{\rm min}$= 0.005 $\mu$m and $a_{\rm max}$=0.1 $\mu$m for graphite.  However in our models  larger grains of $a_{\rm max}$ = 2 $\mu$m are needed in order to accurately fit both the SED and the visibility curves.  Tests were also performed to see how only large grains ($a_{\rm min}$ = 0.2 $\mu$m and $a_{\rm max}$ = 2.0 $\mu$m) would affect the {\tt DUSTY} fitting.
Using $a_{\rm min}$= 0.005 $\mu$m, and $a_{\rm max}$ = 2 $\mu$m provided the best results. While we could achieve adequate fits using only large grains ($a_{\rm min}$= 0.2 $\mu$m and $a_{\rm max}$=2 $\mu$m), the slopes of the visibility curves could not be matched as accurately for RY Sgr and V854 Cen without smaller grains.  Without the smaller grains the screening effect (discussed in \S\ref{ssec:DustyFit}) cannot occur.    %But using only large grains has a significant affect on the output result, $R_{in}$.  Smaller grains are more efficient at absorbing heat, so for similar temperature at the inner boundary, $T_{in}$, smaller dust grains have to be much farther away than larger ones.
Additionally, using only larger grains caused $R_{inner}$ to decrease by a factor of two due to the efficiency of heat absorption.   Such a strong dependence of $R_{inner}$ on the dust grain size leaves this parameter poorly constrained.
It is unclear what the larger dust grains mean in physical terms.  It is unlikely that the dust grains are actually this large.  {\tt DUSTY} may be requiring large dust grains because there is a heavy level of asymmetry which it cannot account for.   More tests need to be done to test all the effects that changing grain size has on {\tt DUSTY} fits of RCB stars.

%It is important to note that the RY Sgr 2005 dataset could only be fit with larger dust grains. This explains why the modeled temperature at the inner radius is slightly higher for the 2005 dataset even though the inner radius is also larger. The larger dust grains used to fit the 2005 data are not as efficient in heating up as the smaller dust grains used to fit the 2007 data.  

\subsection{Future Observations}
\label{ssec:Future}

MIDI's spatial resolution is about 10 mas.  This resolution is not  sufficient to accurately detect and measure the dust forming regions of RCB stars.  However, the Astronomical Multiple BEam combineR (AMBER), a VLTI instrument which observes in the near-IR, has a spatial resolution of 1 to 2 mas \citep{2007NewAR..51..639W}.  This resolution would be enough to directly measure the inner regions of the circumstellar envelope and track the change in location of individual dust clumps.  To do so, frequent observations are needed  (at least a monthly cadence) using the same set of baselines, and  PAs to guarantee that  the same dust cluster or clump is tracked. Although such an observational campaign would require a significant telescope time allocation, the return would be justified: RCB stars are among a few classes of merged objects. In addition, it is likely that the short RCB phase may guarantee that the merger has just taken place. Studying a merger aftermath environment will lead to tremendous insight into the physics of the merger, with possible insight into Type Ia supernova detonation and gravitational wave physics.

\section*{Acknowledgements}
\label{sec:acknowledge}

We acknowledge with thanks the variable star observations from the AAVSO International Database contributed by observers worldwide and used in this research.  The research leading to these results received funding from the European Community's Seventh Framework Programme under Grant Agreement 226604 and from a Macquarie University Research Excellence Scholarship (SNB).  We also wish to thank the anonymous referee for their valuable comments and suggestions.

%  \begin{figure}
%   \centering
%   \includegraphics[width=7.8cm]{V854Cen_ISO_comp.pdf}
%   \includegraphics[width=7.8cm]{V854Cen_ISO_comp2.pdf}
%   \caption{Left: Comparison between the ISO spectrum of V854 Cen and the MIDI data. The ISO flux is well-above the MIDI flux (grism and prism spectra are shown), indicating a large scale dusty nebula around the source. Right: spectra put to a continuum normalized arbitrary scale, in order to compare the level of the dust features. There is no significant differences between the spectra.}
%      \label{fig_vis1}
%    \end{figure}

\bibliography{RCRBsi_arxivSubmit_1}

\label{lastpage}
\end{document}